\documentclass[prx,amsmath,amssymb,twocolumn,superscriptaddress]{revtex4-2}
\usepackage{graphicx}
\usepackage{placeins}
\usepackage{xcolor}
\usepackage[colorlinks,bookmarks=false,linkcolor=blue,urlcolor=blue, citecolor=blue]{hyperref}
\usepackage{pgf, tikz}
\usetikzlibrary{arrows,shapes,positioning}
\usetikzlibrary{datavisualization}
\usetikzlibrary{datavisualization.formats.functions}
\usetikzlibrary[pgfplots.groupplots, decorations, decorations.pathmorphing, decorations.pathreplacing, decorations.text, decorations.markings, backgrounds, chains, calc, automata, arrows.meta, intersections]
\usepackage{pgfplotstable}
\usepackage{svg}
\pgfplotsset{compat=newest}
\pgfplotstableset{col sep=comma}
\usepackage{csvsimple}
\usepackage{tikzscale}
\newcommand{\ex}[1] {\cdot10^{#1}}

\newcommand{\half}{\frac{1}{2}}

\newcommand{\Der}[2]{\ensuremath{\dfrac{\dD #1}{\dD #2}}}

\newcommand{\D}[2]{\ensuremath{\dfrac{\partial #1}{\partial #2}}}

\def \be {\begin{equation}}
\def \ee {\end{equation}}
\def \bc {\begin{center}}
\def \ec {\end{center}}
\def \bf {\begin{figure}[h!]}
\def \ef {\end{figure}}

\def \dD  {{\rm D}}
\def \v {\mathbf}

\def \epsi {\varepsilon}

\def \vnab {\boldsymbol{\nabla}}
\def \vg  {\boldsymbol}
\def \e {{\hspace{.2em}\rm e}}
\def \l {\left}
\def \r {\right}
\def \bee{\begin{equation*}}
\def \eee{\end{equation*}}
\def \ba {\begin{aligned}}
\def \ea {\end{aligned}}

\def \i{\ensuremath{^{-1}}}
\def \ii{\ensuremath{^{-2}}}

\def \ab {{\alpha\beta}}

\def\g{\gamma}

\def \wab {\omega_{\alpha\beta}}

\def \vab {v_{\alpha\beta}}

\def \vgg {v_{\g\g}}

\def \dab {\delta_{\alpha\beta}}

\def \vb {v_\beta}

\def \pt {\partial_t}
\def \pa {\partial_\alpha}
\def \pb {\partial_\beta}

\def \Pg {{p_\gamma}}

\def\ha{{h_\alpha}}
\def\hb{{h_\beta}}

\def \sdab {\sigma_\ab^\text{d}}
\def \stab {\sigma_\ab^\text{tot}}

\def \seab {\sigma_\ab^\text{e}}

\def\Dmu{\Delta\mu}

\def \zp {\zeta_p}
\def \zr {\zeta_\rho}

\definecolor{Gray}{rgb}{0.5, 0.5, 0.5}
\definecolor{Blue}{rgb}{0.0, 0.0, 1.0}
\definecolor{Olive}{rgb}{0.5, 0.5, 0.0}
\definecolor{Maroon}{rgb}{0.5, 0.0, 0.0}
\definecolor{applegreen}{rgb}{0.55, 0.71, 0.0}
\definecolor{ao}{rgb}{0.0, 0.5, 0.0}
\definecolor{bleudefrance}{rgb}{0.19, 0.55, 0.91}
\definecolor{yecolor}{HTML}{ffc311}
\definecolor{nuccolor}{HTML}{0600ff}
\definecolor{myosincolor}{HTML}{26a22b}
\definecolor{cyan}{rgb}{0.0, 1.0, 1.0}
\definecolor{RdPu1}{HTML}{feebe2}
\definecolor{RdPu2}{HTML}{f768a1}
\definecolor{RdPu3}{HTML}{7a0177}

\begin{document}

\title{Defect states in compressible active polar fluids with turnover}

\author{Ludovic Dumoulin}
\affiliation{Department of Biochemistry, University of Geneva, 1211 Geneva, Switzerland}
\affiliation{Department of Theoretical Physics, University of Geneva, 1211 Geneva, Switzerland}

\author{Carles Blanch-Mercader}
\email{carles.blanch-mercader@curie.fr}
\affiliation{Physico-Chimie Curie, Curie Institute, PSL University, 75005 Paris, France}

\author{Karsten Kruse}
\email{karsten.kruse@unige.ch}
\affiliation{Department of Biochemistry, University of Geneva, 1211 Geneva, Switzerland}
\affiliation{Department of Theoretical Physics, University of Geneva, 1211 Geneva, Switzerland}

\date{\today}

\begin{abstract}
Biological active matter like the cytoskeleton or tissues are characterized by their ability to transform chemical energy into mechanical stress. In addition, it often exhibits orientational order, which is essential for many cellular and morphogenetic processes. 
Experimental evidence suggests that defects in the orientational order field play an important role in organizing active stress. However, defects tend to annihilate unless the material is in a chaotic state or hydrodynamic interactions are suppressed.
Using a hydrodynamic description of compressible active polar fluids, we show that turnover readily leads to a stabilization of defects. Depending on the turnover rate, topological defects arrange in a multitude of different phases, including lattices, active foams, and vortex glasses. Our work suggests that turnover plays a crucial role for organizing biological active matter. 
\end{abstract}

\maketitle

\section{Introduction}
The physics of active matter~\cite{Prost:2015ev} is known to play a key role in various morphogenetic processes that occur in single cells or tissues. Well-known examples include the formation of lamellipodia in migrating single cells~\cite{Verkhovsky:1999vj}, apical constriction during gastrulation of the fruit fly \textit{Drosophila melanogaster}~\cite{Martin:2009du}, or cortical actin flows during head-tail symmetry breaking in the nematode \textit{Caenorhabditis elegans}~\cite{Mayer:2010kt}. These processes are driven by gradients in active stress, which is generated autonomously by the biological systems through the conversion of chemical energy released during the hydrolysis of adenosine-triphosphate (ATP) into mechanical work. How gradients in active stress are organized to result in the observed morphogenetic events remains largely unknown.

On a molecular level, active stress is generated by motor proteins that interact with protein filaments, for example, myosin motors and actin filaments. The stress generated in this way is anisotropic due to the elongated form of the filaments. This molecular anisotropy can translate to macroscopic orientational order when filaments align on cellular scales or when elongated cells align on tissue scales. The corresponding orientational order fields are often nematic~\cite{Sanchez.2012,Duclos:2014bs}, when the arrangement is invariant under inversion of the orientation, or polar~\cite{ndlec1997self,delarue2014stress,jain2020role}, when this invariance is broken. Higher order orientational order is also possible~\cite{ArmengolCollado2023}. Active stress is tightly coupled to the orientational order in a cell or tissue~\cite{Dedenon.2025}. 

In recent years, experimental evidence has accumulated that topological defects play an important part in organizing active stress~\cite{Saw:2017gn,Kawaguchi:2017em,Maroudas-Sacks.2021,Guillamat:2022}. These defects are singularities of the orientational field and are characterized by their topological charge~\cite{deGennes:2002vq}. Defects can themselves be oriented~\cite{Vromans.2016,Tang.2017}, but the interaction of defects is not fully accounted for by local properties~\cite{Pearce.2021}. Topological defects can lead to the accumulation of stress~\cite{Saw:2017gn,Blanch-Mercader.2021lzf}, which could be related to the localization of cell death~\cite{Saw:2017gn}, accumulation~\cite{Kawaguchi:2017em} as well as differentiation~\cite{Guillamat:2022}, but also to morphogenetic processes like the formation of protrusions~\cite{Maroudas-Sacks.2021,Guillamat:2022,Maroudas-Sacks.2024,Ravichandran.2025}. 

In passive materials with orientational order, topological defects tend to spontaneously annihilate, unless their existence is required by the boundary conditions or the system's topology through the Poincar\`e-Hopf theorem. In contrast, active stress can lead to a continuous proliferation of singularities in the orientation field~\cite{Sanchez.2012,Thampi:2013cua,Giomi.2015}. However, this proliferation is related to chaotic dynamics -- or low Reynolds number turbulence~\cite{Alert.2022} -- such that it does not seem to be appropriate for a robust organization of active stress. Mechanisms for stabilizing  topological defects in active matter remain largely elusive. Theoretical studies suggest that anisotropic substrates~\cite{Pearce.2019} and coupling to extrinsic curvature can generate orientational defect alignment~\cite{Pearce.2020}. In turbulent states, defect binding has been reported~\cite{Thijssen.2020}. Finally, substrate friction leads to stable defect lattices~\cite{Oza.2016,Doostmohammadi:2016bd,Thijssen.2020a}

\begin{figure*}
\centering
\includegraphics[width=\linewidth]{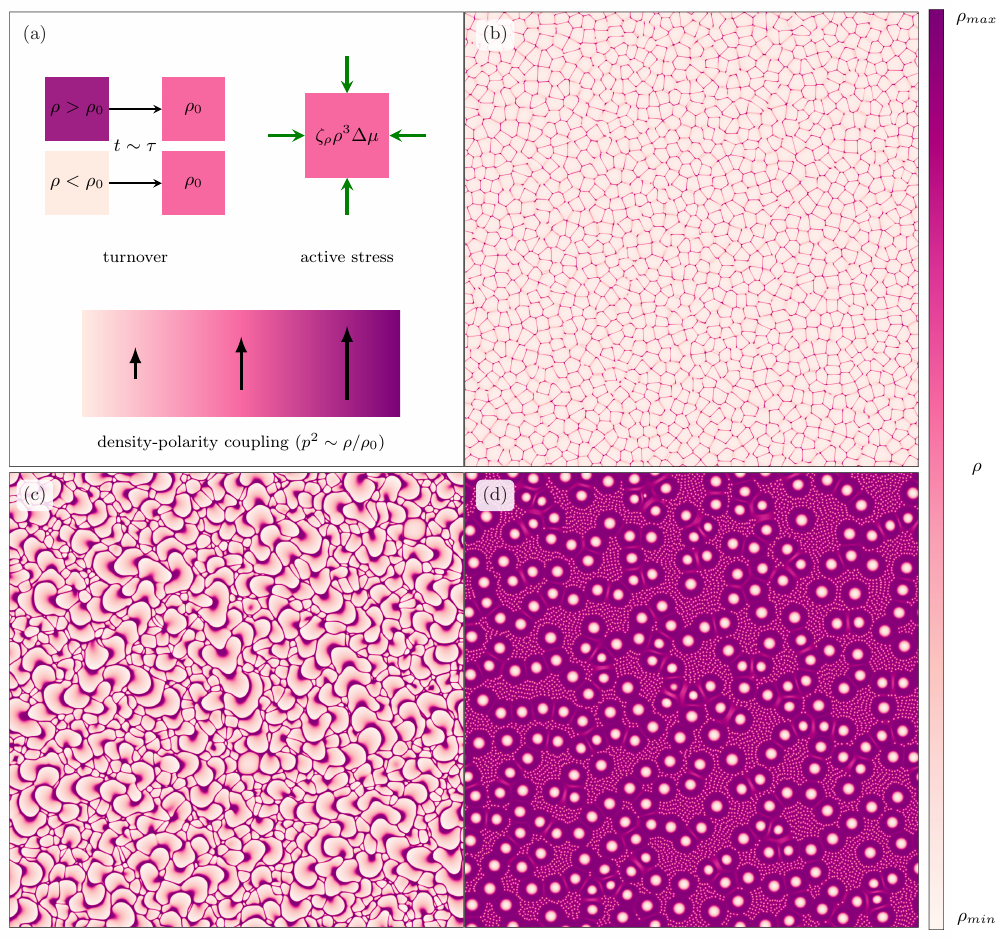}
\caption{Compressible active polar fluids in the presence of turnover. (a) Schematic of the physical processes considered in this work: turnover, active stress, and density-polarity coupling. (b-d) Snapshots of the density heatmaps from numerical solutions of the dynamic equations presented in Sec.~\ref{sec:theory}. Parameters are $\rho_0=0.45$ (b), $\rho_0=0.6$ (c), and $\rho_0=0.75$ (d) as well as $\tau =5$ and $\zeta_\rho=4$. In (b-d),  the system size is $L=50$.  Other parameter values as in Table~\ref{tab:parameterValues}. Maximal and minimal densities are, respectively, $\rho_{max}=0.96$ and $\rho_{min}=0.37$ (b), $\rho_{max}=0.87$ and $\rho_{min}=0.44$ (c), $\rho_{max}=0.79$ and $\rho_{min}=0.58$ (d).} 
\label{fig:schema}
\end{figure*}

In spite of a large body of theoretical work on the dynamics of defects~\cite{shankar2022topological} and the accumulating evidence that topological defects organize active stress during developmental processes~\cite{saw2018biological,maroudas2021mechanical}, the effects of turnover on defects in active fluids are hardly known. In this work, we introduce generalized hydrodynamic equations for compressible active polar fluids in the presence of turnover, Fig.~\ref{fig:schema}(a). We find that turnover can stabilize topological defects. Using long-time numerical solutions of the dynamic equations in large systems, we find a number of different states, including active foams, density waves, and vortex glass states, Fig.~\ref{fig:schema}(b-d), Movies 1-3. In the presence of anisotropic active stress additional phases appear. We show that the linear dynamics of density and polarity perturbations rationalize selection of the phases. These results indicate a central role of turnover in stabilizing topological defects and suggest a basic mechanism for a well-controlled organization of active stress.

\section{Hydrodynamics of an active polar fluid in the presence of turnover}
\label{sec:theory}

Motivated by the actin cortex, we study a two-component system consisting of a compressible active polar fluid and a passive fluid on a two-dimensional space~\cite{Joanny2007,CallanJones2011,Joanny2013,Adar2022}, Fig.~\ref{fig:schema}(a). In the context of the actin cortex, the former represents the actomyosin network and the latter the cytoplasm containing actin monomers and unbound molecular motors. In the following, we derive hydrodynamic equations that describe the system's behavior on large space and time scales. As in Ref.~\cite{Joanny2013}, we neglect permeation flows of the passive component and focus only on the active component from here on. We denote the corresponding mass density by $\rho$ and its velocity field by $\mathbf{v}$. In addition, the active component can exhibit macroscopic polar order, which is captured by the polarity field $\mathbf{p}$. 

The equilibrium behaviour of the system is determined by the total free-energy $\mathcal{F}=\int f d\mathbf{r}$, where $d\mathbf{r}$ is an infinitesimal area element and $f$ is the free-energy density $f$ given by
\begin{align}
f =& \frac{a}{4}\rho^4 +\rho^2\l(-\frac{\chi}{2}\frac{\rho}{\rho_0}|\vg p|^2+\frac{\chi}{4}|\vg p|^4+ \frac{\kappa}{2}(\vnab\vg p)^2\r).\label{eq:freeenergy}
\end{align}
The first term captures the energetic cost for concentrating the active component with an elastic coefficient $a$. We chose this term to be quartic in the mass density $\rho$ to prevent density divergences in the active case. As long as it is large enough, the precise value of the exponent is not important, though. The remaining terms correspond to the Landau-Ginzburg free-energy for the polarity field $\mathbf{p}$~\cite{de1993physics}. For $\chi>0$, we have $|\mathbf{p}|^2 = \rho/\rho_0$ in equilibrium, reflecting an increase of polar order with increasing density. For the time being, $\rho_0$ is some reference density. The final term penalizes distortions of the polarity field with an elastic constant $\kappa$~\cite{de1993physics}. We neglect a potential spontaneous splay term.

In the presence of polar order, the hydrostatic pressure generalizes to the Ericksen stress $\mathsf{\sigma}^\mathrm{e}$ with components
\begin{align}
    \seab &= (f-\rho\mu)\delta_\ab-\D{f}{(\pb p_\gamma)}\pa p_\gamma,\label{eq:ericksenstress}
\end{align}
where greek indices run from 1 to 2 and $\delta_{\alpha\beta}=1$ if $\alpha=\beta$ and zero otherwise. We adopt the convention that summation is implied over repeated indices. In the above expression, $\mu= \frac{\delta \mathcal{F}}{\delta \rho}$ is the active fluid's chemical potential and $\ha = -\frac{\delta \mathcal{F}}{\delta p_\alpha}$ are the components of the molecular field. We give their expressions in App.~\ref{sec:molecular}. The isotropic part of $\mathsf{\sigma}^\mathrm{e}$ can be interpreted as a pressure, $\Pi^\mathrm{eff}=-f+\rho\mu$ with
\begin{align}
    \Pi^\mathrm{eff} = \frac34 a\rho^4-\rho^2\l(\chi\frac\rho{\rho_0}|\mathbf{p}|^2-\frac{\chi}4|\mathbf{p}|^4-\frac{\kappa}{2}(\vnab\vg p)^2\r). \label{eq:pressure}
\end{align}
In equilibrium, when $|\mathbf{p}|^2 = \rho/\rho_0$, the effective isothermal compressibility $\frac{1}{\rho}\frac{\partial\rho}{\partial\Pi}$ is positive only if $\rho_0>\sqrt{5\chi/6a}$. For smaller values of $\rho_0$, the system is thermodynamically unstable.

We now turn to the dynamics of the system. We assume that it is maintained out of equilibrium by the continuous consumption of a chemical fuel. In the case of the actin cortex, this fuel is adenosine tri-phosphate (ATP), which is hydrolyzed to form adenosine di-phosphate (ADP) and inorganic phosphate (P$_i$). The hydrodynamic description is based on conservation laws for mass and momentum and on the broken continuous symmetry under rotations resulting from the polar field~\cite{Kruse:2004il}. 

Mass conservation of the active fluid reads
\begin{align}
    \pt\rho +\pa (\rho v_\alpha)+\pa j_\alpha & = -\tau^{-1}(\rho-\rho_0).
    \label{eq:drhodt}
\end{align}
The total material current is decomposed into a convective part $\rho\mathbf{v}$ and a diffusive part $\mathbf{j}$. The source terms account for assembly and disassembly of the active fluid. We restrict ourselves to small deviations from the target density $\rho_0$ and denote by $\tau$ the characteristic time scale of the disassembly process. The values of $\rho_0$ and $\tau$ are related to the fluid's assembly and disassembly rates $k_a$ and $k_d$ via $\tau=k_d^{-1}$ and $\rho_0=k_a/k_d$. In the context of tissues, these rates are associated with cell division and cell death processes, \cite{shraiman2005mechanical,basan2009homeostatic}. The densities of ATP, ADP, and P$_i$ are not explicitly included in this description, as they are assumed to be spatially and temporally constant.

As cytoskeletal flows or tissue flows are typically associated with a low Reynolds number, we neglect inertial effects. From now on, $\rho$ denotes the number density of the active fluid, which is related to its mass density by some molecular mass that we consider to be constant. Furthermore, momentum conservation reduces to the force balance equation
\begin{align}
\label{eq:divSigma}
    \pb\sigma^\mathrm{tot}_{\alpha\beta} &=\xi v_\alpha
\end{align}
where $\mathsf{\sigma}^\mathrm{tot}$ denotes the total stress tensor. The right hand side corresponds to a viscous friction term with friction coefficient $\xi$. For the actomyosin cortex, it mainly results from interactions between the actomyosin network and the cell membrane or an underlying substrate although permeation of the actomyosin network by the cytosol also contributes to this term. 

To close the system of equations, we need to provide constitutive equations of the active fluid. Following the standard approach of nonequilibrium thermodynamics~\cite{deGroot:1985ue}, we identify the thermodynamic forces conjugate to the fluxes and express the latter to linear order in terms of the former. Explicitly, the fluxes are the symmetric part of the deviatory stress $\mathsf{\sigma}^\mathrm{d}\equiv \mathsf{\sigma}^{\mathrm{tot},s}-\mathsf{\sigma}^{\mathrm{e},s}$, the diffusive current $\mathbf{j}$, the convected co-rotational derivative of the polar field $\mathrm{D}\mathbf{p}/\mathrm{D}t$, and the rate $r$ of ATP hydrolysis. Here, the superscript $s$ denotes the symmetric part of the respective tensors and the convected co-rotational derivative is defined by $\mathrm{D}p_\alpha/\mathrm{D}t = \pt p_\alpha+\vb\pb p_\alpha+\wab p_\beta$, where $\omega_{\alpha\beta}=(\partial_\alpha v_\beta-\partial_\beta v_\alpha)/2$ are the components of the vorticity tensor. Note that the total stress tensor in Eq.~\eqref{eq:divSigma} also contains an anti-symmetric part such that
\begin{align}
\label{eq:sigmaTot}
    \sigma^\mathrm{tot}_{\alpha\beta} &=\sigma_{\alpha\beta}^d+\seab+\frac{1}{2}(p_\alpha h_\beta-p_\beta h_\alpha).
\end{align}

The conjugated thermodynamic forces are the symmetric part of the velocity gradient tensor with components $v_{\alpha\beta}=(\partial_\alpha v_\beta+\partial_\beta v_\alpha)/2$, the gradient of the active fluid's chemical potential $\mu$, the molecular field $\mathbf{h}$, and the difference in chemical potentials of ATP and its hydrolysis products $\Delta\mu=\mu_\mathrm{ATP}-\mu_\mathrm{ADP}-\mu_\mathrm{P}$. Explicitly, the constitutive equations are
\begin{align}
\sdab  & = 2\eta\vab + \frac{\nu}{2}(\ha p_\beta+\hb p_\alpha) \nonumber \\ 
& \quad\quad+\rho\Dmu\l(\zr\rho^2\dab+\zp p_\alpha p_\beta \r)\label{eq:sdab}\\
j_\alpha & = -\gamma\pa\mu \label{eq:ja}\\
\Der{p_\alpha}{t}  & = -\nu p_\beta\vab +\frac{1}{\Gamma}\ha. 
\label{eq:dpdt}
\end{align}
Here, for simplicity, among others we have neglected a term proportional to $\mathbf{p}$ in Eq.~\eqref{eq:ja} and terms containing gradients of the chemical potential $\mu$ in Eqs.~\eqref{eq:sdab} and \eqref{eq:dpdt}.

The constitutive equations contain a number of phenomenological parameters that characterize the material properties. The shear viscosity $\eta$, the rotational viscosity $\Gamma$, the mobility coefficient $\gamma$, and the flow-alignment coefficient $\nu$  are familiar from conventional liquid crystals. In addition, there are active coefficients $\zr$ and $\zp$ that couple the deviatoric stress $\mathsf{\sigma}^\mathrm{d}$ to the chemical potential difference $\Delta\mu$. We have neglected a potential coupling between $\Delta\mu$ and the time derivative of the polarization as it amounts to a redefinition of the equilibrium polarization amplitude 
\cite{kruse2005generic}. In agreement with the Curie principle, we have used the polarization $\mathbf{p}$ to couple fluxes and forces of different tensorial order. Furthermore, we included a dependence on the density $\rho$ in the active terms, because active stress is expected to increase with the active fluid's density. Similar hypotheses have been used in other descriptions of the acto-myosin cortex~\cite{Joanny2013,Levernier2020}. 

In the following, we use the dynamic equations in non-dimensionalized form. Unless stated otherwise we use the parameter values given in Table~\ref{tab:parameterValues}. The values of $\tau$, $\rho_0$, and $\zeta_\rho$ are varied. Additionally, in Eq.~\eqref{eq:sdab} we ignored an isotropic viscous stress $\bar{\eta}\vgg\dab$, an isotropic active stress $\bar{\zp}\Pg\Pg\dab$, and an isotropic elastic stress $\bar{\nu}h_\gamma\Pg\dab$, as we checked that our results are robust to variations in the parameters $\bar{\zp}$ and $\bar{\nu}$. Finally, we set $\Delta\mu=1$ to simplify the notation.

\begin{table*}
\begin{tabular}{c||c|c|c|c|c|c|c|c|c|c|c}
parameter &\quad  $\rho_0'$ \quad & \quad$a'$ \quad & \quad $\chi'$ \quad & \quad $\kappa'$ \quad & \quad $\nu'$ \quad & \quad $\zeta_\rho'$ \quad & \quad $\zeta_p'$ \quad & \quad $\gamma'$ \quad &  $\Gamma'$ \quad & \quad $\tau'$ \quad & \quad $\Dmu$ \quad \\
\hline
value &  \quad 0.4-1.5 \quad & \quad $4/3\zr' \,^*$ \quad & \quad 0.1 \quad & \quad $10^{-4}$ \quad & \quad 0\quad & \quad 0-24 \quad & \quad 0 \quad & \quad $10^{-4}$ \quad & \quad 1 \quad & \quad 0.01-100 \quad & \quad 1 \quad
\end{tabular}
\caption{\label{tab:parameterValues}Values of the dimensionless parameters used in this work. Dimensionless parameters are denoted by the same symbol as the corresponding dimensional parameter, with a prime. The relation between dimensionless and dimensional parameters, as well as the choice of unit system, is detailed in Appendix~\ref{app:dimensionless}. $^*\, a'=4/3$ if $\zr=0$. For simplicity, we omit the prime notation in Secs.~\ref{sec:defects}-\ref{sec:aniso} and in the figures of the main text.}
\end{table*} 
\section{Defect stabilization through turnover}
\label{sec:defects}

\begin{figure}[!t]
\centering
\includegraphics[width=\linewidth]{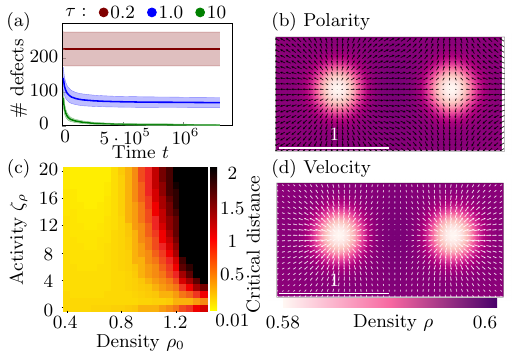}
\caption{(a) Number of defects as function of time for different values of $\tau$ and  $\rho_0 = 1$, $\zeta_\rho = 4$, $L=10$. Solid lines: number of defects averaged over $50$ runs per $\tau$, shaded area: standard deviation. (b) Snapshot of density (heatmap) and polarity (black arrows) maps for a pair of topological defects. Parameters: $\rho_0=0.6, \tau=1, \zr=12$. The image was cropped from a square domain of length $L=10$. (c) Critical distance below which a pair of oppositely-charged defects annihilates, as function of $\rho_0$ and $\zr$ for $\tau=1$ and $L=10$. The black region corresponds to a critical distance larger than half of the system size $L/2$. (d) Snapshot of density (heatmap) and velocity (white arrows) maps for the same defect pair shown in panel (b).}
\label{fig:stab_defects}
\end{figure}

Through minimization of the free energy associated with orientational order, pairs of topological defects with opposite charges tend to merge and annihilate upon relaxation to thermodynamic equilibrium~\cite{Vromans.2016,Tang.2017,Pearce.2021}. In the absence of noise, this relaxation process stops, when the defect configuration with minimal energy and consistent with the boundary conditions is reached. In contrast, active processes can promote the separation of defect pairs and even increase their number~\cite{Sanchez.2012,Giomi:2013ky}. For active nematics, this requires for instance an anisotropic active stress~\cite{Giomi:2014ha}. In polar fluids, we find that turnover can stabilize topological defect pairs both in the presence or absence of active stress, Fig.~\ref{fig:stab_defects}(a). In this section, we analyze the mechanism leading to defect pair separation. To this end, we first consider an isolated defect and then turn to the case of two defects. 

\subsection{An isolated defect}

Our numerical results indicate that the fluid density increases with the distance from a defect center, Fig.~\ref{fig:schema}(b,d). To approximate this continuous variation, we separate space into two domains. Placing the defect center at the origin $r=0$, we define an inner circular domain with a radial distance $r<R$, where we take $\mathbf{p}\equiv \mathbf{p}_\mathrm{in}$ with $|\mathbf{p}_\mathrm{in}|^2=0$ and $\rho=\rho_\mathrm{in}=const$. This inner domain represents the defect core and $R$ represents the radius of a defect core. In the outer domain with $r>R$, we take $|\mathbf{p}|^2\equiv |\mathbf{p}_\mathrm{out}|^2$ and $\rho=\rho_\mathrm{out}=const$. Since the outer domain is much larger than the inner domain, we assume that the density there is set by turnover and choose $\rho_\mathrm{out}=\rho_0$. In addition, we consider that the polar order parameter attains its equilibrium value, such that $|\mathbf{p}_\mathrm{out}|^2=1$. Here, for simplicity, we have chosen to neglect the effect of the Frank free energy, \textit{i.e.}, $\kappa=0$. As a consequence, the Ericksen stress \eqref{eq:sigmaTot} reduces to the pressure \eqref{eq:pressure}. \textit{A priori}, the value of $R$ is determined by the steady state polarity profile. However, since we neglect the Frank free energy, $R$ remains undetermined and is treated as a parameter in the following. In addition, as a consequence of this approximation, we cannot enforce chemical and mechanical balance simultaneously.

If there is a difference between $\rho_\mathrm{in}$ and $\rho_\mathrm{out}$,  there can be a current across the interface between the two domains. This current is driven by differences in the stress and the chemical potential across the domain boundary. Given that the mobility coefficient $\gamma=10^{-4}$ is small in our computations, we neglect the difference in chemical potential and consider only the stress difference as a driver of the flow. Such a flow can be sustained in the presence of turnover.  

The advective flux across the interface $r=R$ is given by $v(R^-)\rho_\mathrm{in}$, where $v(R^-)$ represents the radial component of the fluid velocity in the inner domain evaluated near the interface. To estimate this velocity, we approximate the fluid flow as a Darcy flow, which is proportional to the total stress difference across the interface, $v(R^-)=\Delta\sigma/(\ell\xi)$ with $\Delta\sigma =\sigma(R^+)-\sigma(R^-)$, where $\sigma(R^\pm)$ denote the total stress in the inner (-) and outer domains (+) evaluated near the interface. The parameter $\ell$ is the thickness of the interface, which we assume to be proportional to $R$. Combining Eqs.~\eqref{eq:ericksenstress}, \eqref{eq:pressure}, \eqref{eq:sigmaTot} and \eqref{eq:sdab}, the total stress difference $\Delta\sigma$ is given explicitly by 
\begin{align}
\Delta\sigma &= \frac34 \chi\rho_0^2 - \frac34 a(\rho_0^4-\rho_\mathrm{in}^4) + \zr(\rho_0^3-\rho_\mathrm{in}^3).\label{eq:diffpressure}
\end{align}
Note that mechanical equilibrium $\Delta\sigma=0$ imposes a density difference  $\rho_\mathrm{in}\neq\rho_0$. To make further progress, we assume the density inside the defect to be $\rho_{in} = \rho_0(1-\epsi)$, where $\epsi\ll1$, and expand \eqref{eq:diffpressure} up to linear order with respect to $\epsi$,
\begin{align}
\Delta\sigma&= \frac34 \chi\rho_0^2 -3(a\rho_0- \zr)\rho_0^3\varepsilon + \mathcal O(\varepsilon^2)  \label{eq:pieff}
\end{align}
To determine the value of $\rho_\mathrm{in}$, we impose mass conservation: the rate at which the active fluid is produced in the inner region must equal the rate at which it flows out through the interface. Explicitly, total mass balance condition is given by
\begin{align} 
\tau^{-1}(\rho_0-\rho_\mathrm{in})\pi R^2 &= 2\pi Rv(r=R^-)\rho_\mathrm{in}.\label{eq:massConservation}
\end{align}
Up to first order in $\epsi$, we then get from Eq.~\eqref{eq:massConservation} the fluid velocity 
\begin{align}
v(R^-) &= \frac{\tau^{-1}R}{2}\epsi,\label{eq:advective flux}
\end{align}
where $\epsi$ takes the form
\begin{align}
\epsi &= \frac{3\chi\rho_0^2}{3\chi\rho_0^2+ 12\rho_0^3(a\rho_0-\zr) + 2\xi\tau\i R\ell } .\label{eq:epsi}
\end{align}
Equation \eqref{eq:advective flux} shows that radial flows can arise as an interplay between the density variations that are generated near topological defects and turnover.

Finally to check, whether the result above is consistent with our approximation to neglect the diffusive flux across the interface, we estimate the diffusive flux $j=-\gamma(\mu_\mathrm{out} -\mu_\mathrm{in})/\ell$ and find up to lowest order in $\epsi$
\begin{align}
j&=\gamma\rho_0\chi/\ell.\label{eq:jestimate}
\end{align}
For the advective current we get up to the same order
\begin{align}
v(R^-)\rho_\mathrm{in} & = 3\chi\rho_0^2/(4\xi\ell),\label{eq:flowVelocity}\\
\intertext{such that for the parameters in Table~\ref{tab:parameterValues}, we have the ratio between these two fluxes is}
|j/(v(R^-)\rho_\mathrm{in}) |&= 4\xi\gamma/(3\rho_0)\approx10^{-4}\ll1.
\end{align}

\subsection{A defect pair}

In the case of two defects, the fluid outflow generated near the core of one defect can advect the other and \textit{vice versa}, which tends to separate the two defects.  On the contrary, elastic interactions between the polar fields of the two defects tend to attract defects. The scale of these interaction forces in two dimensions is $\sim\kappa/d$, \cite{deGennes:2002vq}, where $d$ is the distance between the defect centers. Assuming an overdamped dynamics for the defect positions with an effective friction coefficient $\bar{\gamma}$, \cite{Giomi:2014ha,shankar2018defect}, one gets by balancing the above effects 
\begin{align}
\dot{d}=2v(r=d) - \frac{\kappa}{d\bar{\gamma}}.
\end{align}
where $v(r=d)$ represents the velocity field generated by a single defect at position $r=d$. 

For small $d\ll R$, elastic interactions dominate, causing the defects to eventually merge. For large $d\gg R$, elastic interactions dominate again, because they are long-ranged, whereas the radial flows are localized near the defect center. However, at intermediate distances $d\sim R$, radial flows can overcome the elastic attraction, leading to defect separation. This occurs when $\dot{d}>0$ at $d\sim R$, which implies that $\tau^{-1}R^2 \epsi > \kappa/\bar{\gamma}$, where $\epsi$ is given by Eq.~\eqref{eq:epsi} and we used Eq.~\eqref{eq:advective flux} to estimate the fluid velocity. 

\begin{figure}[!ht]
\centering
\includegraphics[width=\linewidth]{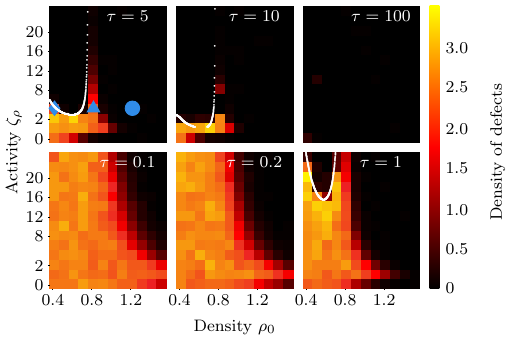}
\caption{Defect density in the $(\rho_0, \zeta_\rho)$-plane for different values of $\tau$. For each data point, a single numerical solution for a system size of $L=10$ has been analyzed. White dots indicate the critical activity $\zeta_\rho^c$ obtained by solving Eq.~\eqref{eq:zetaRhoDeltaMuC} with $a=4\zeta_\rho^c/3$ (values of $\zeta^c_\rho<1$ are not shown). Blue symbols indicate solutions for $\rho_0=0.4$ (diamond), $0.8$ (triangle), and $1.2$ (circle) shown in Fig.~\ref{fig:states} below. Other parameter values as in Table~\ref{tab:parameterValues}.}
\label{fig:Ndefects}
\end{figure}

This expectation is confirmed by numerical integration of the dynamics of two oppositely charged defects, which qualitative agrees with the analytic considerations, Fig.~\ref{fig:stab_defects}(b-d), Movies 4 and 5. Numerical results show that turnover of the active fluid can stabilize topological defects, Fig.~\ref{fig:stab_defects}(c). In large systems, this leads to a sustained finite density of defects, Fig.~\ref{fig:Ndefects}, maintained up to critical values of the density $\rho_0$ and the activity $\zeta_\rho$. This is compatible with our simplified analysis above: as $\rho_0$ and $\zeta_\rho$ increase, the value of $\varepsilon$ \eqref{eq:epsi} decreases, leading to a reduction in the amplitude of the radial flows \eqref{eq:advective flux}, and thus repulsive interaction between defects. Furthermore, the energetic cost of a defect increases with increasing the target density $\rho_0$, promoting defect annihilation. Finally, the defect density decreases with $\tau$, Fig.~\ref{fig:stab_defects}(a) and Fig.~\ref{fig:Ndefects}, showing that sufficiently fast turnover is necessary for defect stabilization, which is again consistent with our simplified analysis. 

\section{Defect states for isotropic active stress}
\label{sec:defectStates}

In this section, we show how topological defects arrange as a function of the target density $\rho_0$, which is set by the turnover rates. We start with a linear stability analysis of the homogenous polarized state. Then, we present numerical solutions of the dynamic equations, and finally rationalize the observed sequence of states using a linear stability analysis of the isotropic state.

\subsection{Linear stability analysis}
\label{sec:linStab}

\begin{figure}
\centering
\includegraphics[width=\linewidth]{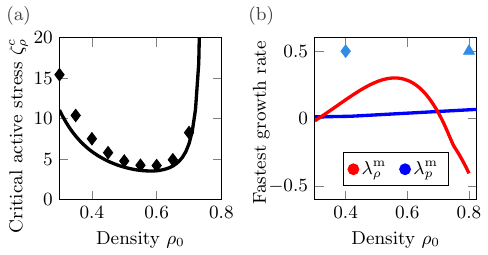}
\caption{Linear stability analysis of homogenous states. (a) Stability diagram of the homogenous polar stationary state as a function of $\rho_0$ and $\zeta_\rho$ for $\tau=5$ and $a=4/3\zr^c$. Solid lines: density and polarity perturbations, diamonds: only density perturbations. (b) Fastest growth rate of density (red) and polarity (blue) perturbations of the homogenous isotropic state for $\zeta_\rho=4$ and $\tau=5$, other parameters as in Table~\ref{tab:parameterValues}.}
\label{fig:linstab}
\end{figure}

Consider the uniform polarized state, $\rho = \rho_0$ and $\mathbf{p}=\mathbf{p}_0$ with $|\mathbf{p}_0|^2=1$, which is a steady state of the dynamic equations \eqref{eq:freeenergy}-\eqref{eq:dpdt}. The linearized dynamics for a small perturbation of this state is given in App.~\ref{app:linStab}. The ensuing stability diagram shows that this state is unstable for a critical level of active stresses $\zeta_\rho>\zeta_{\rho}^{c}$, where the critical value $\zeta_{\rho}^{c}$ is a nonmonotonic function of the target density $\rho_0$, Fig.~\ref{fig:linstab}(a). 

The instability is mainly driven by fluctuations in the density. To see this, we computed the growth rate associated with density fluctuations $\lambda_{\rho}$ in the special case $\chi=0$, which takes the form
\begin{equation}
\lambda_\rho(q)=-\frac{(q^2-q_c^2)^2}{\tau q_c^4(1+2q^2)}+\frac{3q^2\rho_0^3(\zeta_\rho-\zeta_\rho^c)}{1+2q^2}.
\end{equation}
The variable $q_c$ denotes the wavenumber of the perturbation with the maximal growth rate $\lambda_\rho$ at the critical value $\zeta_\rho=\zeta_\rho^c$ and is given by
\begin{equation}
q_c=\frac{1}{(2\gamma A(\chi=0)\tau)^{1/4}}.
\end{equation}
where $A=3a\rho_0^2-5\chi/2$, see App.~\ref{app:linStabPolarized}.

The variable $\zeta_\rho^c$ is the critical level of the active stress at which the growth rate at $q=q_c$ vanishes, i.e. $\lambda_\rho(q_c,\zeta_\rho=\zeta_\rho^c)=0$, and is given by
\begin{equation}
\zeta_\rho^c = \frac{A(\chi=0)}{3\rho_0}+\left(\sqrt{\frac{A(\chi=0)\gamma}{3\rho_0^3}}+\sqrt{\frac{2}{3\rho_0^3\tau}}\right)^2\label{eq:zetaRhoDeltaMuC}
\end{equation}
This critical activity threshold provides a good approximation for the value $\zeta^c_\rho$, when the value of $\chi=0.1$, Fig.~\ref{fig:linstab}(a). 

Let us point out that the linear dynamics are qualitatively different in absence of turnover. In that case, density perturbations with $q=0$ are marginal, implying a type II instability according to the classification of Cross and Hohenberg~\cite{Cross:1993el}. In contrast, the instability in the presence of turnover is of type I.

\subsection{Nonlinear dynamics}

To explore the system dynamics beyond linear stability, we numerically solved the dynamic equations \eqref{eq:freeenergy}-\eqref{eq:dpdt} using a custom code written in Julia~\cite{bezanson2017julia}. Details are given in App.~\ref{app:numerical}. The code can be found at~\cite{github}. Typically, we used square domains of size $L=10$ with a discretization length $\Delta x=10^{-2}$ and a total simulated time of $10^5$. Selected simulations were run for a total simulated time of $2\ex6$ or square domains of size $L=50$. Units are set as described in App.~\ref{app:dimensionless}.

\begin{figure*}
\centering
\includegraphics[width=\linewidth]{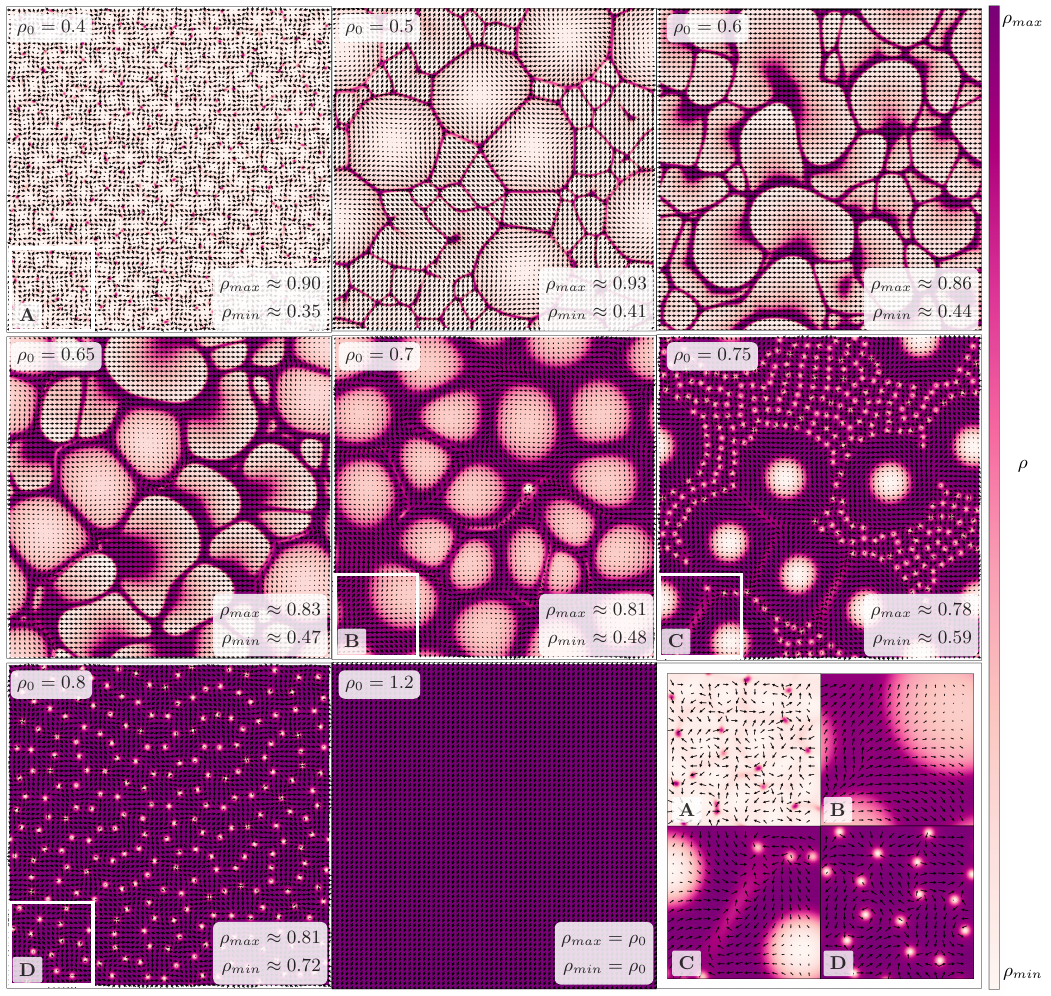}
\caption{Snapshots of the density (heatmaps) and polarity (black arrows) of asymptotic states of the dynamic equations \eqref{eq:freeenergy}-\eqref{eq:dpdt} for different target densities $\rho_0$ for $\tau=5$ and $\zeta_\rho = 4$. The system size is $L=10$. Other parameters as in Table~\ref{tab:parameterValues}. The same states are presented for a system of length $L=50$ in Fig.~\ref{fig:statesL50}. The bottom right panel shows zoomed-in views of selected snapshots, labeled A–D. Maximal and minimal densities are indicated in legends.}
\label{fig:states}
\end{figure*}

Simulations are initiated with a slight perturbation of the uniform disordered state by randomly perturbing both the uniform density, $\rho=\rho_0$, and the uniform polarity field, $\mathbf{p}=\mathbf{0}$. We observed a variety of different asymptotic behaviors. In the following, we describe the different phases observed in the computations, Fig.~\ref{fig:states}.

\subsubsection{Puncta and networks} 
For small target densities $\rho_0$, $\rho_0=0.4$ in Fig.~\ref{fig:states} and Movie 6, the system spontaneously forms puncta of elevated density within a background with a density around $\rho_0$. The majority of these puncta are connected to other puncta by lines of intermediate density that formed the edges of a network with puncta at its vertices. We denote the regions enclosed by edges as cells. 

The network is dynamic and undergoes topological changes as puncta move, fuse, and split. Simultaneously, polar order is present for these parameter values, Fig.~\ref{fig:states}A. The polarity does not exhibit a preferred orientation at the cell edges. Furthermore, the polar field exhibits defects with integer charges $\pm1$. Defect centers are mainly located within the cells. 

\subsubsection{Active foams} 
With increasing target density $\rho_0$, the density along the edges increases, $\rho_0=0.5$ in Fig.~\ref{fig:states} and Movie 7. The system forms two sorts of cells: One kind contains a single integer topological defect of the polar order field, whereas cells of the other kind do not contain defects. In the latter, the polarity is typically aligned throughout the whole cell. Some edges have a vertex that is not connected to other edges. This typically occurs in cells without a defect in the polar order field. Similar to the case of $\rho_0=0.4$, this state is dynamic and we denote it in the following as an active foam. We discuss similarities and differences with  active foam states reported in other systems in Sect.~\ref{sec:discussion}.

\subsubsection{Density waves} 
\label{sec:densityWaves} 
As $\rho_0$ is increased further, traveling density waves appear, $\rho_0=0.6$ and $\rho_0=0.65$ in Fig.~\ref{fig:states} and Movies 2 and 8. At the same time, uniform polar order emerges throughout the whole system. In contrast to lower densities, the cell shapes are elongated in one direction. The direction of elongation is on average perpendicular to the direction of motion of the density waves, but not correlated with the direction of the polarity field. These states are identical to the waves obtained when starting with a fully polarized state $\mathbf{p}=\mathbf{p}_0=const$ with $|\mathbf{p}_0|^2=1$. At $\rho_0=0.7$, Fig.~\ref{fig:states} and Movie 9, we observe re-entrant behavior of active foams. This time the interfaces are thicker along the edges compared to active foams at lower densities, Fig.~\ref{fig:states}B. This state also features traveling density waves.

\subsubsection{Vortex glass} 
For a target density of $\rho_0=0.75$, topological defects with cores of two different sizes emerge, Fig.~\ref{fig:states} and Movie 3. The state is dynamic with ``small'' defects flowing between ``large'' ones. Due to the similarity of this state with states observed in the complex Ginzburg-Landau equation, we denote this state as a vortex glass state~\cite{Brito.2003}. Note, however, that the disclination points with charge +1 can also be asters and spirals in addition to vortices, Fig.~\ref{fig:states}C.

\subsubsection{Defect lattices} 
As the target density is increased further, the size of defect cores is again monodisperse. For densities around $\rho_0=0.8$, locally, regular defect lattices appear spontaneously, Fig.~\ref{fig:states}D and Movie 10. The topological defects have charges of $\pm1$ and the lattices are either square or hexagonal. We discuss the lattice phase in more detail in Sect.~\ref{sec:defectLattices} below.

\subsubsection{Homogenous uniformly polarized state} 
For target densities $\rho_0$ beyond a critical value, the system asymptotically settles into the homogenous state with uniform polarity, Movie 11. Indeed, the critical value $\zeta_\rho^c$, Eq.~\eqref{eq:zetaRhoDeltaMuC}, increases with the density $\rho_0$. By increasing $\rho_0$ and keeping all other parameter values constant, the homogenous uniformly polarized state becomes linearly stable. Our numerical results suggest that for sufficiently large densities $\rho_0$ the state is globally stable except for a uniform rotation of the polar field.

\subsection{Competition between density and polarity perturbations}
\label{sec:competition}

We can rationalize the transitions between the states in Fig.~\eqref{fig:states} as a function of the average density $\rho_0$ by comparing the rates associated with the fastest growing mode of density fluctuations $\lambda_{\rho}^m$ and polarity fluctuations $\lambda_{p}^m$ around the isotropic homogeneous state. The calculation of these two rates is given in App.~\ref{app:linStab}. For the parameters used in Fig.~\ref{fig:states}, we present their values as a function of $\rho_0$ in Fig.~\ref{fig:linstab}(b).

For intermediate target densities $\rho_0\gtrsim 0.35$ and $\rho_0\lesssim 0.7$, density fluctuations grow faster than polarity fluctuations, Fig.~\ref{fig:linstab}(b). Correspondingly, density patterns dominate in this parameter regime and impose polarity patterns, Fig.~\ref{fig:states}. With increasing average density $\rho_0$, the extent of the high-density domains increases. 

Because the growth rate $\lambda_{p}^m$ grows quadratically with $\rho_0$, whereas the growth rate $\lambda_{\rho}^m$ is non-monotonic with $\rho_0$, there is a cross-over at about $\rho_0\simeq 0.75$ to a regime where polarity fluctuations grow faster than density fluctuations, Fig.~\ref{fig:linstab}(b). In this case, the final pattern results from the interaction between defects, see states at $\rho_0=0.75$ and $\rho_0=0.8$ in Fig.~\ref{fig:states}. At even larger target densities $\rho_0>0.85$, the system always relaxes to a polarized homogeneous phase, Fig.~\ref{fig:states}. 

\section{Defect lattices}
\label{sec:defectLattices}

\begin{figure}
\begin{center}
\includegraphics[width=\linewidth]{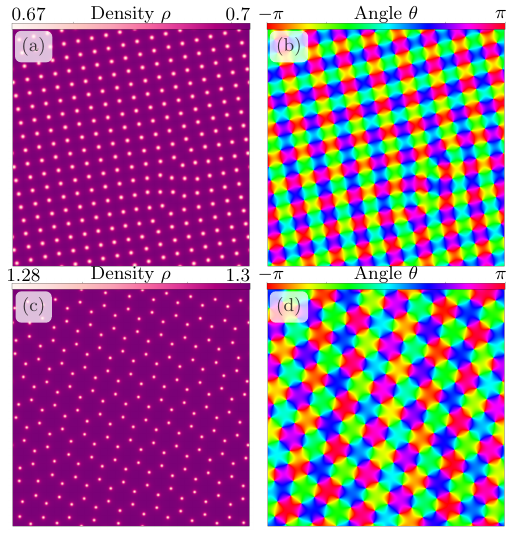}
\end{center}
\caption{Examples of square and hexagonal defect lattices. (a,c) Density heatmaps at $t=2\ex6$ for $\rho_0 = 0.7$, $\tau = 1$, and $\zeta_\rho=10$ (a) and for $\rho_0 = 1.3$, $\tau = 0.2$, and $\zeta_\rho=1$ (c). (b,d) Angle of the polarity field corresponding to (a) and (c). System size $L=10$, other parameters as in Table~\ref{tab:parameterValues}. The corresponding time evolution of the shape order $\Gamma_p$ in Eq.~\eqref{eq:ShapFunction} is given in Fig.~\ref{fig:lattice1extended}} \label{fig:lattice1}.
\end{figure}

In the defect lattices we described in the previous section, the topological defects in the polarity field locally arrange into regular polygons. This behavior is robust across varying parameter values. To further characterize these states, we perfromed longer simulations with a total time of $t=2\ex{6}$. Over time, a global crystalline order of defect positions emerged,  Fig.~\ref{fig:lattice1}. We observed both, square and hexagonal lattices,  Fig.~\ref{fig:lattice1}(a,c).

Let us recall that numerical computations start from a slightly perturbed uniform disordered state. Due to local spontaneous symmetry breaking, defects in the polarity field appear throughout the system, initially forming an irregular arrangement. The defects created in this way are stabilized by the mechanism discussed in Sect.~\ref{sec:defects} and eventually arrange in a nearly disclination-free lattice as in Fig.~\ref{fig:lattice1}(a,c).

A striking feature of these lattices is that they move at constant velocity without defect rearrangements, Movies 12 and 13. Defect lattices move at constant velocity along a direction that is not determined by the symmetry axes. 

A detailed inspection reveals an alternating pattern of defects with topological charges $+1$ and $-1$, Figs.~\ref{fig:lattice1}(b,d) and \ref{fig:lattice1extended}(a,d). In the one-constant approximation, isolated $+1$ defects of either type - aster, spiral, and vortex - have the same total free-energy, \cite{deGennes:2002vq}. These subtypes appear in both square and hexagonal defect lattices, but do not show any specific spatial organization.

The formation of defect lattices with clearly peaked structure factors takes too much time as if a thorough characterization of the corresponding phase space were feasible. To quantify the emerging crystalline order from shorter computer runs, we use the fact that the emergence of long-range translational order in the defect arrangement is preceded by long-range orientational order and use a method described in Ref.~\cite{ArmengolCollado2023} for its quantification. We start by computing the Voronoi tesselation of defect centers, which assigns a polygonal cell to each defect based on its surrounding defects. From these polygons, we compute the shape function $\gamma_p$ for $p=1,2,3,\ldots$, which quantifies the degree of orientational order in the spatial arrangement of defect centers. Explicitly, for a single cell
\be
\gamma_p = \frac{1}{\Delta_p}\sum_{v=1}^V|\v r_v|^p\e^{ip\phi_v},
\ee
where $V$ is the number vertices of the cell, $\mathbf{r}_v$ the position of vertex $v$ relative to the defect center, $\phi_v$ its angle with the $x$-axis, and $\Delta_p = \sum_{v=1}^V|\v r_v|^p$. For a regular $p$-polygon, one has $\gamma_p=1$, because  $|\mathbf{r}_v|=const$ and because the angle between two adjacent vertices, $\measuredangle(\mathrm{r}_v,\mathrm{r}_{v+1})$, is $2\pi/p$. Finally, for each $p$, we average the absolute value of the shape function $\gamma_p$ over all defect centers, 
\begin{equation}
\Gamma_p=\langle|\gamma_p|\rangle.\label{eq:ShapFunction} 
\end{equation}
To assess the degree of order, we compare $\Gamma_p$ to $\Gamma_p^0$, its value for randomly distributed vertices, by computing the difference $\Gamma_p - \Gamma_p^0$.

\begin{figure}
    \centering
    \includegraphics[width=\linewidth]{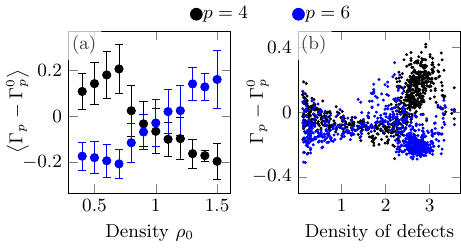}
    \caption{Transition from square to hexagonal defect lattices. (a) Average of the shape function $\langle\Gamma_p-\Gamma_p^0\rangle$ given by Eq.~\eqref{eq:ShapFunction}. For each value of $\rho_0$, the average is taken with respect to numerical solutions with $\tau=0.1$, 0.2, 0.5, 1, 1.2, 1.7, 2.5, 5, 10, 100 and $\zeta_\rho= 0$, 1, 2, 4, 6,\ldots, 24. Only solutions with defect density larger than $1.5$ were analyzed. (b) Shape function $\Gamma_p-\Gamma_p^0$  as a function of the defect density for the solutions used in (a). Error bars in panel (a) are SD. In both panels, the domain size $L=10$.}
    \label{fig:gamma_rho}
\end{figure}

The trajectories of $\Gamma_p-\Gamma_p^0$, which we computed for $p=2,\ldots,6$, indicate early on, $t\lesssim10^5$, the developing crystalline order present in the arrangement of defects, Fig.~\ref{fig:lattice1extended}(c,f). For each parameter set that displays a defect lattice, we use $\Gamma_p$ at $t=1.4\cdot10^5$ to characterize the organization of the defects. Although, in principle, this quantity can depend on the initial conditions, we find that it essentially depends on the defect density Fig.~\ref{fig:gamma_rho}b.  Specifically, at sufficiently small defect densities, the dominant type of order is hexatic ($p=6$). In contrast, at intermediate defect densities, the dominant order changes to quartic symmetry  ($p=4$), indicating  that the emergence of quartic crystalline order is not an effect of packing constraints. This observation is further supported by defects with smaller cores arranging in triangular lattices, while those with bigger cores form square lattices. At intermediate defect densities ($>1.5$ per unit area), the crystalline order in the spatial arrangement of defect centers can also be controlled by the target density $\rho_0$, Fig.~\ref{fig:gamma_rho}a.

\section{Effects of dynamic anisotropies} 
\label{sec:aniso}

In the previous section, we studied the effect of turnover in the absence of anisotropic active stresses and flow alignment. In the following, we show that flow alignment affects the size of defect cores and can induce a transition to the vortex glass state. Furthermore, we show the emergence of labyrinthine patterns in the presence of an anisotropic active stress. 

\subsection{Flow alignment}

The flow-alignment parameter $\nu$ controls the coupling between the dynamics of the polarity field and shear flows. In addition to affecting the magnitude of polarity, the value of $\nu$ also determines how shear flows reorient the polarity field. 

To gain insight into the role of $\nu$, we first review a simpler case:  a polarization field subjected to a linear shear flow of the form $\mathbf{v} = \gamma(-x, y)$, where $x$ and $y$ are cartesian coordinates and $\gamma>0$ denotes the shear strength~\cite{deGennes:2002vq}. In this setting, compression occurs along the $x$-direction, while expansion occurs along the $y$-direction. For $\nu<-1$, which corresponds to rod-shaped conventional liquid crystals, the polarity tends to align with the expansive direction. In contrast, for $\nu>1$, which is characteristic of disc-shaped liquid crystals, the polarity tends to align with the compressive direction. 

In our computations, we observed that single topological defects can create an outward radial velocity field that  decays with the distance from the defect center, as discussed in Section~\ref{sec:defects}. Consequently, compression occurs in the radial direction. Based on the reasoning above, we expect flow-alignment to favor vortex-type defects for $\nu<-1$, and aster-type defects for $\nu>1$.

\begin{figure}[ht]
\centering
\includegraphics[width=\linewidth]{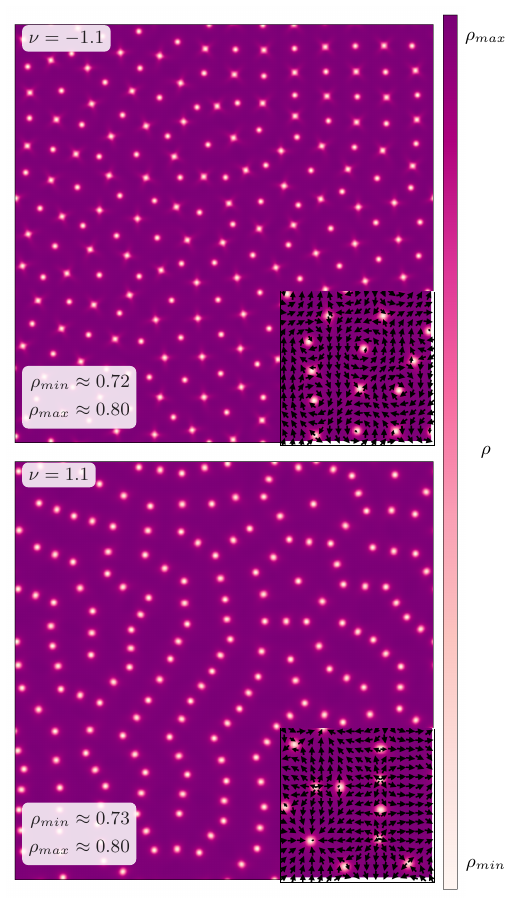}
\caption{Effect of flow-alignment on +1 topological defects. Density (heatmap)and polarity (black arrows) for $\rho_0 = 0.8$, $\tau = 0.2$, and $\zeta_\rho=4$. Changing the flow alignment parameter $(\nu)$ can turn all the +1 topological defects into vortices $(\nu=-1.1)$ or asters ($\nu=1.1$). In both cases, the fluid has not yet reached a steady state. The domain size is $L=10$.}
\label{fig:flowalign}
\end{figure}

Next, we studied the effect of the flow-alignment parameter on defect lattices. Previously, for $\nu=0$, we observed an alternating pattern of $+1$ and $-1$ defects, where the $+1$ defects could be of any type, aster, spiral, or vortex, Fig.~\ref{fig:lattice1extended}. For $\nu\neq0$, in contrast, flow alignment leads to a preference of asters for $\nu>0$ and of vortices for $\nu<0$, Fig.~\ref{fig:flowalign}. This is in agreement with our simplified analysis above. To assure continuity of the polarization field, the direction of the polarization field is opposite in neighboring asters or neighboring vortices, see arrows in insets in Fig.~\ref{fig:flowalign}.

Furthermore, changing the flow alignment parameter affects the density and distribution of defects. This is most readily observed at low target densities $\rho_0$, Figs.~\ref{fig:flowalign} and \ref{fig:flowalignandDensity}. The defect density is larger for $\nu<0$ than for $\nu>0$, Fig.~\ref{fig:flowalignandDensity}. Also, it increases with $\nu$. At these intermediate stages of the solutions, the defects form strings. Furthermore, the density around a defect is anisotropic, see for example Fig. \ref{fig:flowalignandDensity} case $\nu=0.5$ and $\rho_0=0.75$. 

\subsection{Anisotropic active stress}

Next we explore the effects of an anisotropic active stress $\zeta_p\neq0$ on the long-term system dynamics. Anisotropic active stress is known to qualitatively change the dynamics of an active polar fluid. For example, they lead to sustained spontaneous flows~\cite{Kruse:2004il,Voituriez:2007jy,Duclos:2018ita,Dedenon.2023,Chandragiri.2020}.

\begin{figure}[t]
\centering
\includegraphics[width=\linewidth]{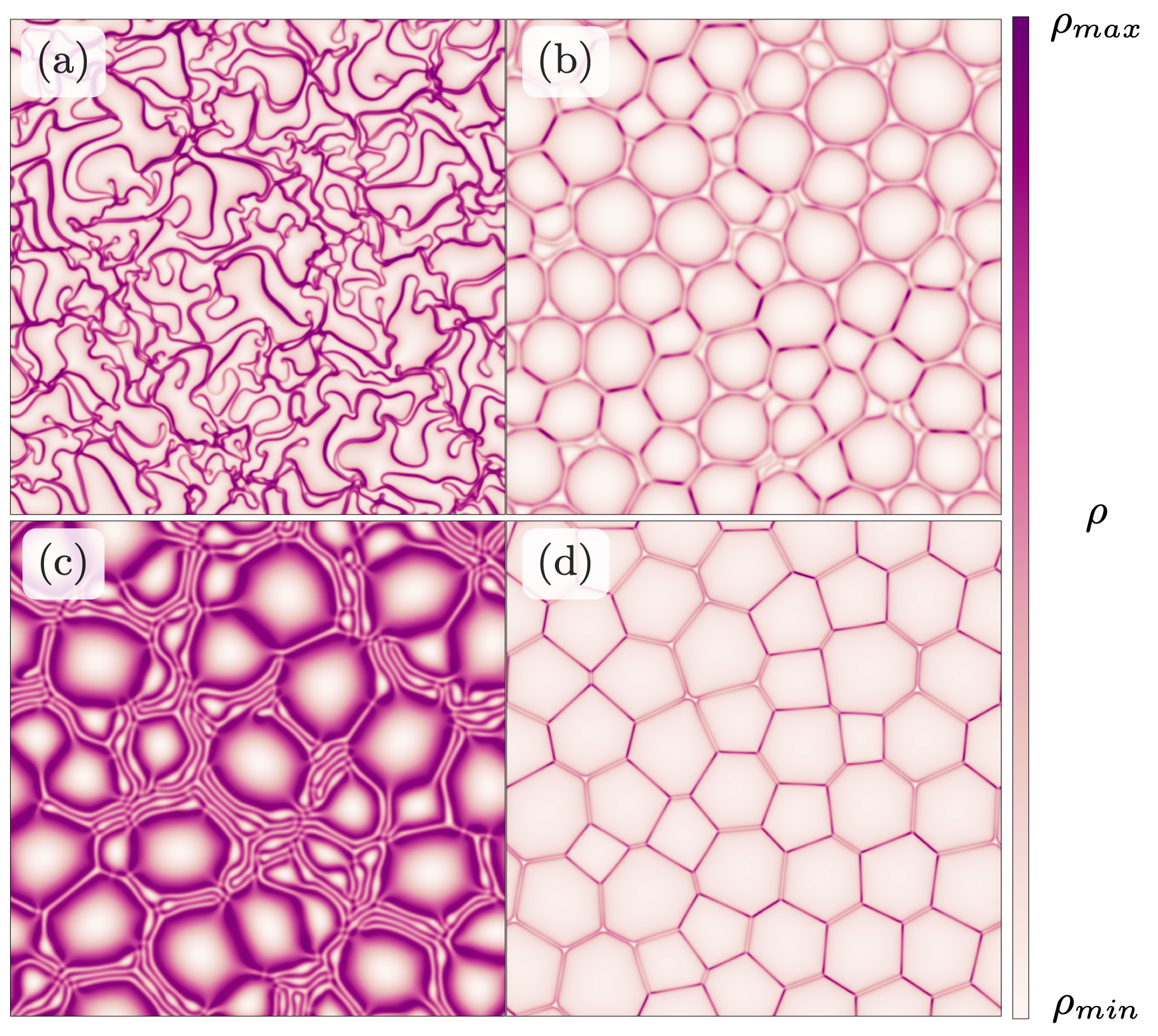}
\caption{Effects of active anisotropic active stress. (a) Contractile anisotropic stress without flow alignment. Parameters: $\rho_0 = 0.5$, $\tau = 4.0$, $\zeta_\rho=4$, $\nu = 0$ and $\zeta_p=-1$. (b) Extensile anisotropic stress with negative flow alignment. Parameters: $\rho_0 = 0.4$, $\tau = 4$, $\zeta_\rho=4$, $\nu = -0.5$ and $\zeta_p=1$. (c) Extensile anisotropic stress with positive flow alignment. Parameters: $\rho_0 = 0.6$, $\tau = 1$, $\zeta_\rho=8$, $\nu = 0.5$ and $\zeta_p=0.5$. (d) Extensile anisotropic stress without flow alignment. Parameters: $\rho_0 = 0.4$, $\tau = 2$, $\zeta_\rho=8$, $\nu = 0$ and $\zeta_p=1$. The domain size is $L=10$. Other parameter values as  in Table~\ref{tab:parameterValues}.}
\label{fig:anisotropicStress}
\end{figure}

In Figure~\ref{fig:anisotropicStress}(a) and Movie 14, we present a snapshot of the system for $\zeta_p=-1$. We observe the emergence of a labyrinthine density pattern. The paths of the labyrinth can be closed, for example circular, or exhibit bifurcations. In regions free of bifurcations, the paths tend to align. 
Labyrinthine patterns have also been found in defect free active nematic fluids~\cite{Lavi.2024}. In contrast, to the patterns reported here, the labyrinths are formed by nematic domain walls, whereas the density is constant.

In Figure~\ref{fig:anisotropicStress}(b-d) and Movies 15-17, we present snapshots of solutions for extensile anisotropic active stress. In contrast to Fig.~\ref{fig:anisotropicStress}(a), we also consider flow alignment. For negative flow alignment, $\nu<0$, the fluid forms high density rings, some of which are connected by bridges, Fig.~\ref{fig:anisotropicStress}(b). Typically, the density is higher along segments that are not adjacent to segments of other rings. For positive flow alignment, $\nu>0$, there are again large regions of low density, but they are limited by broader high-density regions, Fig.~\ref{fig:anisotropicStress}(c). Besides, there are several lines of high density between the low-density regions. In the absence of flow alignment, $\nu=0$, but still extensile anisotropic active stress, Fig.~\ref{fig:anisotropicStress}(d), we observe again an active foam state. Some adjacent foam bubbles have separate surfaces. All the structures shown in Fig.~\ref{fig:anisotropicStress} are dynamic.

These examples of solutions in the presence of flow alignment and anisotropic active stress component show that they can either have rather subtle effects by affecting the type of defect or more dramatic consequences by leading to features not observed in their absence, namely, disordered high-density lines.

\section{Discussion}
\label{sec:discussion}

In this manuscript, we studied the effect of turnover on the dynamics of active polar fluids with a focus on couplings between the density and polarity fields. Through a combination of extensive numerical computations and analytical calculations, we found various macroscopic phases, most notably active foams and defect lattices. The former are characterized by low-density regions surrounded by high-density interfaces, forming a dynamic network that continuously re-arranges. The latter corresponds to arrays of integer topological defects that over time develop long-ranged crystal order, in particular triangular and square lattices. The topological defects are stabilized by an interplay between density gradients at the defect core and turnover. This mechanism is independent of the strength of the anisotropic active stress, unlike the mechanism for defect pair separation in active nematic fluids, \cite{Giomi:2014ha,shankar2018defect}. Finally, we discussed two extensions of the theoretical description to account for alignment by shear flows and anisotropic active stress. 

Theoretically, a contractile network with some similarity to the active foam discussed above has been obtained in Ref.~\cite{Staddon.2022}. In that work, turnover of an active fluid is assumed to be regulated by a biochemical reaction network. However, in that description, the active fluid is isotropic and the contractile network might approach a stationary state. Finally, an active foam state has been reported for so-called Malthusian flocks in the limit of constant density~\cite{Besse.2022}. There, the "bubbles" of the foam are delimited by line defects in the orientational order field. 

Experimentally, an active foam state has been reported in three-dimensional microtubule-motor mixtures~\cite{Lemma.2022tcq}. Motor accumulation at microtubule ends is an important feature of this state. As our theory does not distinguish between filaments and motors, it is not obvious how to relate it to the experiments on motor-microtubule networks. Furthermore, in the experiment, the accumulation of motors at microtubule ends rather leads to high motor-filament densities in asters, which is opposite to the depletion of active fluid generated by the free energy density \eqref{eq:freeenergy}. We conclude that the mechanisms underlying active foam formation are different in the experiment~\cite{Lemma.2022tcq} and our theory. More recently, cytoplasmic compartmentalization by microtubule asters has been reported in zebrafish embryos and frog extracts~\cite{Rinaldin.2024}. Phenomenologically, these states resemble the active foam state shown in Fig.~\ref{fig:states} for $\rho=0.7$. Whether our description appropriately captures the mechanism underlying compartmentalization in these systems remains to be seen.

There have been a number of theoretical works on defect lattices in active fluids~\cite{Doostmohammadi:2016bd,Oza.2016,Thijssen.2020a,James.2021,Pearce.20218dr,Keogh.2022,Caballero.2023}. In two-dimensional nematic active fluids, defect lattices were notably reported in the case of high surface friction compared to internal friction~\cite{Doostmohammadi:2016bd,Thijssen.2020a,Caballero.2023}. In these works the active anisotropic stress was extensile. As we have shown in Sect.~\ref{sec:defects}, in our polar fluid, turnover leads to a repulsive interaction between defects in presence of contractile activity. At intermediate values of the target density, this repulsion can lead to the formation of defect lattices.

Turnover is a general feature of living matter, on subcellular as well as tissue scales. We thus expect our findings to be relevant in various biological situations. It should be noted, though, that in the absence of fluctuations, as it was considered in this work, topological defects are not created by the dynamics of the contractile active fluid. Instead, they are generated through the initial condition. To robustly yield a certain density of defects, turnover could be regulated biochemically. In future work, it will be interesting to study the effects of couplings between biochemical regulation and the dynamics of active fluids.  

\appendix

\section{Expressions for the chemical potential, the molecular field and the Ericksen stress} \label{sec:molecular}

In this section, we give the explicit expressions of the chemical potential $\mu=\frac{\delta \mathcal{F}}{\delta\rho}$, the molecular field $\mathbf{h}=-\frac{\delta \mathcal{F}}{\delta\mathbf{p}}$, and the Ericksen stress $\mathsf{\sigma}^\mathrm{e}_{\alpha\beta}$ given by Eq.~\eqref{eq:ericksenstress}. The total free-energy is $\mathcal{F}=\int f d\mathbf{r}$, where $d\mathbf{r}$ is an infinitesimal area element and $f$ is the free-energy density given by Eq.~\eqref{eq:freeenergy}. We obtain
\begin{align}
\mu =& a\rho^3 + \rho\l(-\frac{3}{2}\chi\frac{\rho}{\rho_0} p_\gamma^2+\frac{1}{2}\chi(p_\gamma^2)^2+\kappa(\partial_\gamma p_\delta)^2 \r)\label{eq:mu} \\
\ha =& \chi\rho^2\l(\frac{\rho}{\rho_0} -  p_\gamma^2\r)p_\alpha + \kappa\pb(\rho^2\pb p_\alpha)\label{eq:ha}\\
\sigma^\mathrm{e}_{\alpha\beta} =& -\left[\frac{3}{4}a\rho^4-\rho^2\left(\chi\frac{\rho}{\rho_0}p_\gamma^2-\frac{1}{4}\chi(p_\gamma^2)^2\right)\right]\dab\nonumber\\&-\rho^2\left[\frac{1}{2}\kappa(\partial_\gamma p_\delta)^2 \right]\dab -\kappa\rho^2(\partial_\alpha p_\gamma)(\partial_\beta p_\gamma).
\end{align} 

\section{Dimensionless equations}\label{app:dimensionless}
\def \taut {\tilde\tau}
\def \rhot {\tilde\rho}

Let us begin by defining the units used in our theoretical description. The time unit is set by an arbitrary time scale $T$, the length scale by $\lambda = \sqrt{\eta\xi\i}$ and the viscosity scale by $ \eta$. Combining these expressions, we define a density scale as  $\rhot = \eta T\lambda\ii$. 

In this unit system, the dimensionless parameters are: $a' = a\rhot^4 T\eta\i$, $\chi' = \chi\rhot^2 T\eta\i$, $\kappa'=\kappa \rhot^2 T\eta\i\lambda\ii$, $\rho_0'=\rho_0 \rhot\i$, $\tau'=\tau T\i$,
$\nu'=\nu$, $\zeta_\rho' = \zeta_\rho\rhot^3 T\eta\i$, $\zeta'_p = \zeta_p\rhot T\eta\i$, $\gamma' = \gamma\eta\lambda\ii\rhot\ii$, and $\Gamma'=\Gamma\eta\i$. 

Finally, we set $\Delta\mu=1$ to simplify the notation.

\section{Linear stability analysis of the homogeneous state}
\label{app:linStab}
In this section, we present the linear stability analysis of two homogeneous states: the polarized state and the isotropic state. We focus on the special case with no flow alignment, $\nu=0$ and no anisotropic stress, $\zeta_p=0$. 

\subsection{Polarized homogeneous state}
\label{app:linStabPolarized}

To study the linear stability of the polarized state, we consider without limiting generality that the polar field is aligned along the $x$-axis, $\mathbf{p}_0.\mathbf{e}_y=0$. We set $\rho=\rho_0+\delta\rho$, $\mathbf{p}=\mathbf{p}_0 + \delta\mathbf{p}$ with $|\mathbf{p}_0 |=1$, and $\mathbf{v}=\delta\mathbf{v}$, and derive the dynamic equations to first order in the perturbations $\delta\rho$, $\delta\mathbf{p}$, and $\delta\mathbf{v}$. By setting $\delta\rho=\hat{\rho}(q_x,q_y)\mathrm{e}^{i(q_x x + q_y y)}$ and similarly defining the Fourier components $\hat{v}_x$ and $\hat{v}_y$ for the velocity perturbation $\delta\mathbf{v}$ and $\hat{p}_x$ and $\hat{p}_y$ for the polarity perturbation, we use force balance~\eqref{eq:divSigma} to express $\hat{v}_x$ and $\hat{v}_y$ in terms of $\hat{\rho}$, $\hat{p}_x$ and $\hat{p}_y$. Explicitly,
\begin{align}
\hat{v}_x &= \frac{-iq_x \rho_0}{2q^2+1}\left[\left(A-3\rho_0\zr\right)\hat{\rho} - \chi\rho_0 \hat{p}_x\right] - \half\frac{iq_yq^2\rho_0^2}{q^2+1}\kappa\hat p_y \\
\hat{v}_y &= \frac{-iq_y \rho_0}{2q^2+1}\left[\left(A-3\rho_0\zr\right)\hat{\rho} - \chi\rho_0 \hat{p}_x\right] + \half\frac{iq_xq^2\rho_0^2}{q^2+1}\kappa\hat p_y.
\end{align}
Here, as in the main text, $A=3a\rho_0^2-\frac{5}{2}\chi>0$ and $q^2=q_x^2+q_y^2$ denotes the squared magnitude of the wavevector $\mathbf{q}=(q_x,q_y)$.

Substituting the expressions for $\hat{v}_x$ and $\hat{v}_y$ into the linearized equations~\eqref{eq:freeenergy}-\eqref{eq:dpdt}, we obtain, to linear order in perturbations, the following dynamical equations for the density and the polarization
\begin{align}
\partial_t\hat{\rho} &= -\left[\tau^{-1}+B(q)\right]\hat{\rho} +
C(q)\hat{p}_x\label{eq:c3}\\
\partial_t\hat{p}_x &= \frac{\chi\rho_0}{\Gamma}\hat{\rho} - \rho_0^2\frac{2\chi+\kappa q^2}{\Gamma}\hat{p}_x\label{eq:c4}\\
\partial_t\hat{p}_y &
=-\kappa\rho_0^2q^2\l(\frac{1}{\Gamma}+\frac14\frac{q^2}{q^2+1}\r)\hat{p}_y\label{eq:c5}
\end{align}
with 
$B(q)=q^2\left[\rho_0^2\left(A-3\rho_0\zr\right)/(1+2q^2)+\gamma A\right]$ and 
$C(q) = q^2\chi\rho_0\left[\rho_0^2/(1+2q^2)+\gamma\right]$. Note that perturbations in the polarization component perpendicular to the x-axis $p_y$ always decay, as shown in Eq.~\eqref{eq:c5}. The linear dynamics of perturbations in $\rho$ and $p_x$ are coupled in general, and the resulting expressions for the two associated growth rates are rather involved, such that we do not present them here. However, valuable insights can be gained by considering the special case $\chi=0$, in which the linear dynamics of $\rho$ and $p_x$  decouple. In this case, perturbations in $p_x$ always decay, see Eq.~\eqref{eq:c4}, while perturbations in $\rho$ evolve with a growth rate $\lambda_\rho$, which can be recast in the form
\begin{equation}
\lambda_\rho(q)=-\frac{(q^2-q_c^2)^2}{\tau q_c^4(1+2q^2)}+\frac{3q^2\rho_0^3(\zeta_\rho-\zeta_\rho^c)}{1+2q^2}.
\end{equation}
The variable $q_c$ denotes the wavenumber of the perturbation with the maximal growth rate $\lambda_\rho$ at the critical value $\zeta_\rho=\zeta_\rho^c$ and is given by
\begin{equation}
q_c=\frac{1}{(2\gamma A(\chi=0)\tau)^{1/4}}.
\end{equation}
The variable $\zeta_\rho^c$ is the critical level of the active stress at which the growth rate at $q=q_c$ vanishes, i.e. $\lambda_\rho(q_c,\zeta_\rho=\zeta_\rho^c)=0$, and is given by
\begin{equation}
\zeta_\rho^c = \frac{A(\chi=0)}{3\rho_0}+\left(\sqrt{\frac{A(\chi=0)\gamma}{3\rho_0^3}}+\sqrt{\frac{2}{3\rho_0^3\tau}}\right)^2
\end{equation}

\subsection{Isotropic homogeneous state}
\label{app:linStabIsotropic}

In our numerical integration of the equations ~\eqref{eq:freeenergy}-\eqref{eq:dpdt}, we initialize the system in the isotropic state with a small random perturbation. In Sect.~\ref{sec:competition}, to interpret the  states observed in numerics, we use a linear stability analysis of the isotropic state. Here, we provide a detailed explanation of this analysis. 

We set $\rho=\rho_0+\delta\rho$, $\mathbf{p}=\delta\mathbf{p}$, and $\mathbf{v}=\delta\mathbf{v}$, and derive the dynamic equations to first order in the perturbations $\delta\rho$, $\delta\mathbf{p}$, and $\delta\mathbf{v}$. In Fourier space (see previous section), we use force balance~\eqref{eq:divSigma} to express $\hat{v}_x$ and $\hat{v}_y$ in terms of $\hat{\rho}$, $\hat{p}_x$ and $\hat{p}_y$. Explicitly,
\begin{align}
\hat{v}_x &= 3E\frac{iq_x}{2q^2+1}\hat{\rho}\\
\hat{v}_y &= 3E\frac{iq_y}{2q^2+1}\hat{\rho}
\end{align}
with $q^2=q_x^2+q_y^2$ and 
$E=(\zeta_\rho-a\rho_0)\rho_0^2$. Substituting the expressions above for $\hat{v}_x$ and $\hat{v}_y$ into the linearized equations~\eqref{eq:freeenergy}-\eqref{eq:dpdt}, we obtain the following dynamical equations for the density and the polarization
\begin{align}
\frac{d}{dt}\hat{\rho} &= -\left\{\frac{1}{\tau}+3q^2\left[a\rho_0^2\gamma -\rho_0\frac{E}{2q^2+1}\right]\right\}\hat{\rho}.\label{eq:C2_eq1}\\
\frac{d}{dt}\hat{\mathbf{p}} &=\frac{\rho_0^2}{\Gamma}\left(\chi-\kappa q^2\right)\hat{\mathbf{p}}\label{eq:C2_eq2}.
\end{align}
As long as $\chi>0$, the isotropic state is unstable to perturbations in the polarity field,  with a growth rate increasing quadratically with the target density $\rho_0$. 

In Fig.~\ref{fig:linstab}, we represent the rates associated with the fastest growing mode of density fluctuations $\lambda_{\rho}^m$ and polarity fluctuations $\lambda_{p}^m$ around the isotropic homogeneous state. These rates were calculated using \eqref{eq:C2_eq1} and \eqref{eq:C2_eq2}. For example, the rate $\lambda_{p}^m=\chi\rho_0^2/\Gamma$ is quadratic in the target density $\rho_0^2$.  

\section{Numerical methods} \label{app:numerical}

In this appendix, we provide some details on the numerical used to integrate~\eqref{eq:freeenergy}-\eqref{eq:dpdt}. The code is available at \cite{github}. 

Spatial derivatives were calculated by first transforming the dynamic fields to the reciprocal space using the two-dimensional Fast Fourier Transform (FFT) technique, then multiplication with the corresponding wave vectors and, finally, transforming back to real space using FFT. Temporal integration was executed through the explicit Euler forward method. The computational grid encompassed at least $1008x1008$ discrete points, with a spatial discretization of $\Delta x = 10^{-2}$. The maximal temporal discretization is set to $\Delta t_\text{max} = 10^{-2}$ and $\Delta t$ is chosen at every time unit $\Delta t$ according to the following rule $\Delta t = \text{min}\l\{ \Delta t_\text{max}, 5\ex{-2}\Delta x/\text{max}\l(||v|| \r),1.25\Delta t_\text{prev}\r\}$ with $\Delta t_\text{prev}$ being the previous time step. These computations were performed on GPUs.

At each time step, the chemical potential $\mu$ Eq.~\eqref{eq:mu}, the molecular field $\ha$ Eq.~\eqref{eq:ha}, and the total stress minus the viscous stress $\sigma^{\rm{nv}}_\ab = \stab - 2\eta\vab$, were computed in real space.
Then, the velocity field was computed by solving the force balance Eq.~\eqref{eq:divSigma} in Fourier space. Finally, the temporal the density and polarity fields were updated by time integration of Eqs.~\eqref{eq:drhodt} and~\eqref{eq:dpdt}.

\section{Supplementary figures}

In this Section, we provide additional figures.
\begin{figure*}
    \centering
    \includegraphics[width=\linewidth]{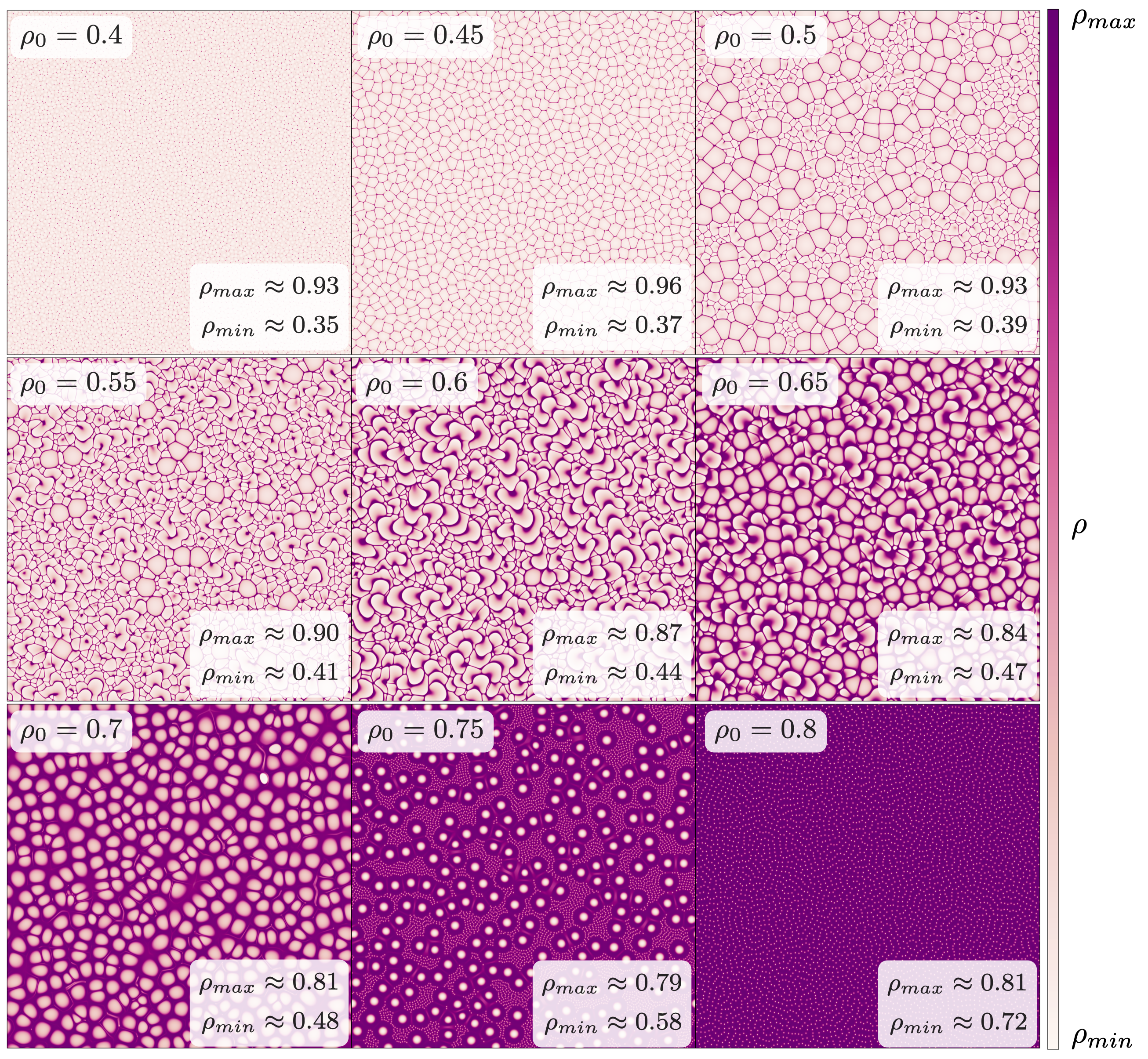}
    \caption{Snapshots of the density (heatmaps)  for different target densities $\rho_0$ and for $L=50$, $\tau=5$ and $\zeta_\rho = 4$. Other parameters as in Table~\ref{tab:parameterValues}. The cases of $\rho_0=0.45$ and $\rho_0=0.55$ are not represented in Fig.~\ref{fig:states}.}
    \label{fig:statesL50}
\end{figure*}

\begin{figure*}
    \bc
    \includegraphics[width=\linewidth]{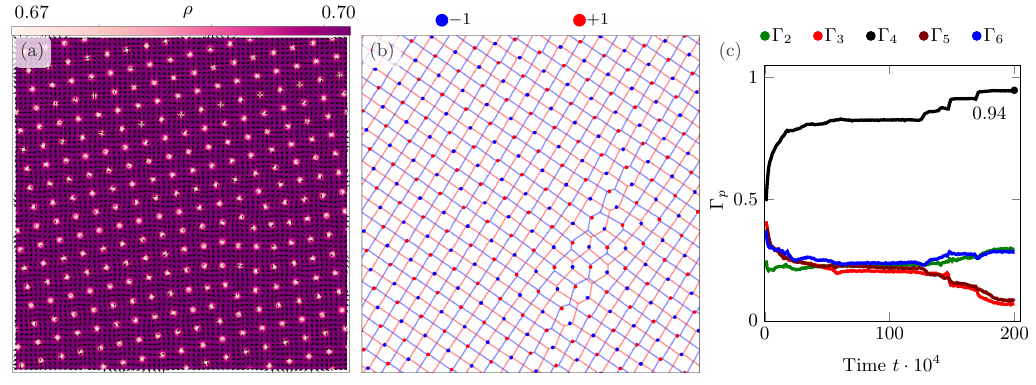}
    \includegraphics[width=\linewidth]{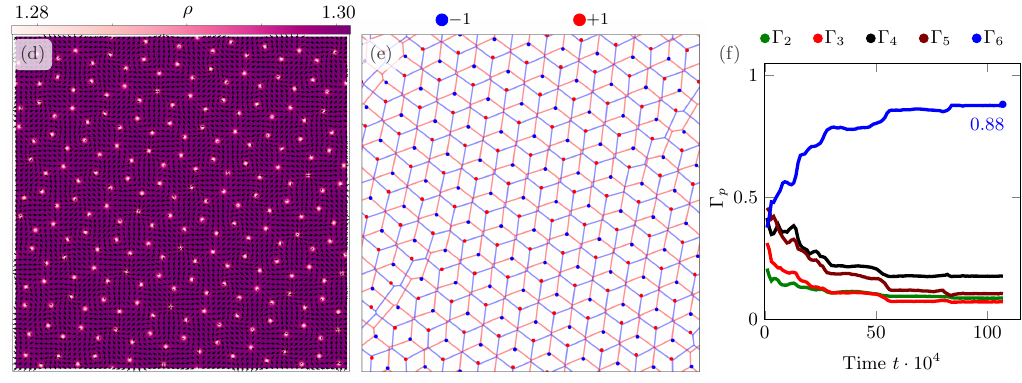}
    \ec
    \caption{Example of a square and hexagonal defect lattice. (a,d) Snapshot at $t=2*10^6$ of the density (heatmaps) and polarity (black arrows). (b,e) Voronoi tessellation for the patterns in (a) and (d). (c,f) Time evolution of the shape order parameter $\Gamma_p$ in Eq.~\eqref{eq:ShapFunction} for $p=2$ to $6$ for the cases in (a) and (d). In panels (a-c) $\rho_0 = 0.7$, $\tau = 1$, and $\zeta_\rho=10$ and in panels (d-f) $\rho_0 = 1.3$, $\tau = 0.2$, and $\zeta_\rho=1$. The system size was set to $L=10$. Other parameters as in Table~\ref{tab:parameterValues}. } \label{fig:lattice1extended}
\end{figure*}

\begin{figure*}
    \centering
    \includegraphics[width=\linewidth]{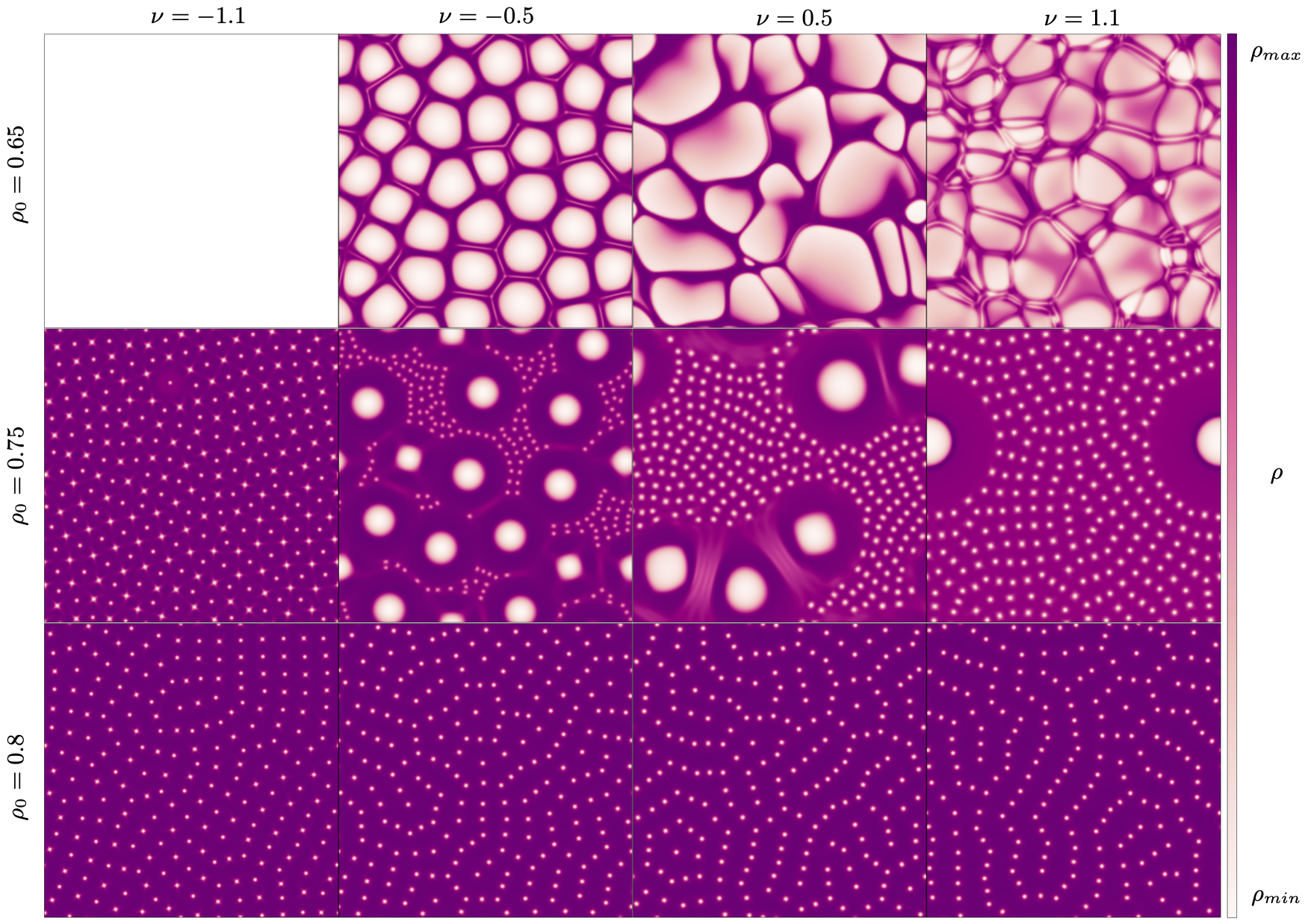}
    \caption{Snapshots of the density (heatmaps) of solutions of the dynamic equations \eqref{eq:freeenergy}-\eqref{eq:dpdt} for different target densities $\rho_0$  and flow alignment parameter $\nu$ and for $\tau=5$ and $\zeta_\rho = 4$. The system size is $L=10$. Other parameters as in Table~\ref{tab:parameterValues}.}
    \label{fig:flowalignandDensity}
\end{figure*}

\FloatBarrier

\section{Movies}

In this section, we provide captions for the Movies.

\textbf{Movie 1.} Active fluid density from a numerical solution of the dynamic Eqs.~\eqref{eq:drhodt}, \eqref{eq:divSigma}, and \eqref{eq:dpdt} corresponding to Fig.~\ref{fig:schema}(b) for $L=50$, $\rho_0=0.45$, $\tau=5$, and $\zeta_\rho=4$. Other parameters as in Table~\ref{tab:parameterValues}. Total time is $28\ex3$. 

\textbf{Movie 2.} Active fluid density from a numerical solution of the dynamic Eqs.~\eqref{eq:drhodt}, \eqref{eq:divSigma}, and \eqref{eq:dpdt} corresponding to Fig.~\ref{fig:schema}(c) for $L=50$, $\rho_0=0.6$, $\tau=5$, and $\zeta_\rho=4$. Other parameters as in Table~\ref{tab:parameterValues}. Total time is $28\ex3$. 

\textbf{Movie 3.} Active fluid density from a numerical solution of the dynamic Eqs.~\eqref{eq:drhodt}, \eqref{eq:divSigma}, and \eqref{eq:dpdt} corresponding to Fig.~\ref{fig:schema}(d) for $L=50$, $\rho_0=0.75$, $\tau=5$, and $\zeta_\rho=4$. Other parameters as in Table~\ref{tab:parameterValues}. Total time is $28\ex3$. 

\textbf{Movie 4.} Active fluid density and polarity from a numerical solution of the dynamic Eqs.~\eqref{eq:drhodt}, \eqref{eq:divSigma}, and \eqref{eq:dpdt} for two defects of opposite charges corresponding to Fig.~\ref{fig:stab_defects}(b), $\rho_0=0.6$, $\tau=1$, and $\zeta_\rho=12$. Other parameters as in Table~\ref{tab:parameterValues}. Total time is $2\ex3$. 

\textbf{Movie 5.} Active fluid density and velocity field from a numerical solution of the dynamic Eqs.~\eqref{eq:drhodt}, \eqref{eq:divSigma}, and \eqref{eq:dpdt} for two defects of opposite charges corresponding to Fig.~\ref{fig:stab_defects}(b), $\rho_0=0.6$, $\tau=1$, and $\zeta_\rho=12$. Other parameters as in Table~\ref{tab:parameterValues}. Total time is $2\ex3$. 

\textbf{Movie 6.} Active fluid density and polarity from a numerical solution of the dynamic Eqs.~\eqref{eq:drhodt}, \eqref{eq:divSigma}, and \eqref{eq:dpdt} for $L=25$, $\rho_0=0.4$, $\tau=5$, and $\zeta_\rho=4$. Other parameters as in Table~\ref{tab:parameterValues}. Total time is $128\ex3$. 

\textbf{Movie 7.} Active fluid density and polarity from a numerical solution of the dynamic Eqs.~\eqref{eq:drhodt}, \eqref{eq:divSigma}, and \eqref{eq:dpdt} for $L=25$, $\rho_0=0.5$, $\tau=5$, and $\zeta_\rho=4$. Other parameters as in Table~\ref{tab:parameterValues}. Total time is $128\ex3$. 


\textbf{Movie 8.} Active fluid density and polarity from a numerical solution of the dynamic Eqs.~\eqref{eq:drhodt}, \eqref{eq:divSigma}, and \eqref{eq:dpdt} for $L=25$, $\rho_0=0.65$, $\tau=5$, and $\zeta_\rho=4$. Other parameters as in Table~\ref{tab:parameterValues}. Total time is $128\ex3$. 

\textbf{Movie 9.} Active fluid density and polarity from a numerical solution of the dynamic Eqs.~\eqref{eq:drhodt}, \eqref{eq:divSigma}, and \eqref{eq:dpdt} for $L=25$, $\rho_0=0.7$, $\tau=5$, and $\zeta_\rho=4$. Other parameters as in Table~\ref{tab:parameterValues}. Total time is $128\ex3$. 


\textbf{Movie 10.} Active fluid density and polarity from a numerical solution of the dynamic Eqs.~\eqref{eq:drhodt}, \eqref{eq:divSigma}, and \eqref{eq:dpdt} for $L=25$, $\rho_0=0.8$, $\tau=5$, and $\zeta_\rho=4$. Other parameters as in Table~\ref{tab:parameterValues}. Total time is $128\ex3$. 

\textbf{Movie 11.} Active fluid density and polarity from a numerical solution of the dynamic Eqs.~\eqref{eq:drhodt}, \eqref{eq:divSigma}, and \eqref{eq:dpdt} for $L=25$, $\rho_0=1$, $\tau=5$, and $\zeta_\rho=4$. Other parameters as in Table~\ref{tab:parameterValues}. Total time is $128\ex3$. 

\textbf{Movie 12.} Active fluid density and polarity from a numerical solution of the dynamic Eqs.~\eqref{eq:drhodt}, \eqref{eq:divSigma}, and \eqref{eq:dpdt} for $L=10$, $\rho_0=0.7$, $\tau=1$, and $\zeta_\rho=10$. Other parameters as in Table~\ref{tab:parameterValues}. Total time is $2\ex6$. 

\textbf{Movie 13.} Active fluid density and polarity from a numerical solution of the dynamic Eqs.~\eqref{eq:drhodt}, \eqref{eq:divSigma}, and \eqref{eq:dpdt} for $L=10$, $\rho_0=1.3$, $\tau=0.2$, and $\zeta_\rho=1$. Other parameters as in Table~\ref{tab:parameterValues}. Total time is $2\ex6$. 

\textbf{Movie 14.} Active fluid density and polarity from a numerical solution of the dynamic Eqs.~\eqref{eq:drhodt}, \eqref{eq:divSigma}, and \eqref{eq:dpdt} for $L=10$, $\rho_0=0.5$, $\tau=4$, $\zeta_\rho=4$, $\nu=0$, and $\zeta_p=-1$. Other parameters as in Table~\ref{tab:parameterValues}. Total time is $40\ex3$. 

\textbf{Movie 15.} Active fluid density and polarity from a numerical solution of the dynamic Eqs.~\eqref{eq:drhodt}, \eqref{eq:divSigma}, and \eqref{eq:dpdt} for $L=10$, $\rho_0=0.4$, $\tau=4$, $\zeta_\rho=4$, $\nu=-0.5$, and $\zeta_p=1$. Other parameters as in Table~\ref{tab:parameterValues}. Total time is $40\ex3$. 

\textbf{Movie 16.} Active fluid density and polarity from a numerical solution of the dynamic Eqs.~\eqref{eq:drhodt}, \eqref{eq:divSigma}, and \eqref{eq:dpdt} for $L=10$, $\rho_0=0.6$, $\tau=1$, $\zeta_\rho=8$, $\nu=0.5$, and $\zeta_p=0.5$. Other parameters as in Table~\ref{tab:parameterValues}. Total time is $40\ex3$. 

\textbf{Movie 17.} Active fluid density and polarity from a numerical solution of the dynamic Eqs.~\eqref{eq:drhodt}, \eqref{eq:divSigma}, and \eqref{eq:dpdt} for $L=10$, $\rho_0=0.4$, $\tau=2$, $\zeta_\rho=8$, $\nu=0$, and $\zeta_p=1$. Other parameters as in Table~\ref{tab:parameterValues}. Total time is $40\ex3$. 

\FloatBarrier

\bibliography{refs}

\begin{thebibliography}{65}
\bibitem [{git()}]{github}%
  \BibitemOpen
  \href@noop {} {}\bibinfo {note} {Code will be made available upon publication.}\BibitemShut {Stop}%
\end{thebibliography}

\end{document}